\documentclass[reprint]{revtex4}
\usepackage{amsfonts,latexsym,eucal,amsthm,amssymb} 
\usepackage{graphicx}
\usepackage{psfrag}
\usepackage{color}
\usepackage{amsbsy}
\usepackage{lineno}
\usepackage[final]{changes}
\colorlet{Changes@Color}{red}
\usepackage{mathtools}
\usepackage{comment}

\def\c{{\bf c}}
\def\dd{\mbox{d}}

\def\ve{\varepsilon}

\def\n{{\bf n}}

\def\x{{\bf x}}

\def\l{\ell}

\usepackage{wrapfig}
\usepackage{dsfont}
\usepackage{enumerate}
\usepackage[sort&compress]{natbib}

\begin{document}

\title{Immigration-induced phase transition in a regulated
  multispecies birth-death process}



\author{Song Xu$^{1}$ and Tom Chou$^{1,2}$ \\
$^{1}$Dept. of Biomathematics, UCLA, Los Angeles, CA 90095-1766 \\
$^{2}$Dept. of Mathematics, UCLA, Los Angeles, CA 90095-1555}


\begin{abstract}
  Power-law-distributed species counts or clone counts arise in many
  biological settings such as multispecies cell populations,
  population genetics, and ecology.  This empirical observation that
  the number of species $c_{k}$ represented by $k$ individuals scales
  as negative powers of $k$ is also supported by a series of
  theoretical birth-death-immigration (BDI) models that consistently
  predict many low-population species, a few intermediate-population
  species, and very high-population species. However, we show how a
  simple global population-dependent regulation in a neutral BDI model
  destroys the power law distributions.  Simulation of the regulated
  BDI model shows a high probability of observing a high-population
  species that dominates the total population.  Further analysis
  reveals that the origin of this breakdown is associated with the
  failure of a mean-field approximation for the expected species
  abundance distribution.  We find an accurate estimate for the
  expected distribution $\langle c_k \rangle$ by mapping the problem
  to a lower-dimensional Moran process, allowing us to also
  straightforwardly calculate the covariances $\langle c_k c_\l
  \rangle$.  Finally, we exploit the concepts associated with energy
  landscapes to explain the failure of the mean-field assumption by
  identifying a phase transition in the quasi-steady-state species
  counts triggered by a decreasing immigration rate.
\end{abstract}

\maketitle

\section{Introduction}

High-dimensional stochastic models are important across many fields of
science and often arise in biological contexts such as T cell receptor (TCR)
diversity in immunology \cite{zarnitsyna2013estimating}, species
abundance and diversity in ecology \cite{mcgill2007species}, and
populations in cellular barcoding experiments
\cite{goyal2015mechanisms}.
T cells in jawed vertebrates can be classified
into multiple subpopulations, each corresponding to different T cell
receptor (TCR) subtypes produced in the thymus.  Here the number of T
cells $n_i$ expressing the $i^{\rm th}$ receptor represents the
$i^{\rm th}$ dimension.  In this setting the large number of different
TCRs ($1\leq i \leq \Omega$, $\Omega \sim 10^6 - 10^{8}$) present in
an organism allows its adaptive immune system to recognize and respond
to a wide range of antigens that it might encounter.  Multispecies
ecological communities are another example of high-dimensional
systems. If the habitat of interest is an
island, then $n_i$ quantifies the number of animals of species $i$ on
the island. The gut is also a habitat for many coexisting species of
  bacteria that make up the microbiome \cite{MICROBIAL_GUT}.
Finally, DNA-tagging and sequencing technology has allowed \textit{in
  vivo} tracking of multiple hematopoietic clones, each of which was
derived from a unique hematopoietic stem cell that carries a unique
DNA tag \cite{zarnitsyna2013estimating, kim2014dynamics,
  sun2014clonal, biasco2016vivo, koelle2017quantitative}, resulting in
clonal-tracking data of very high dimensions \cite{goyal2015mechanisms,SONG_BLOOD}.

\begin{figure}[h!]
  \begin{center}
    \begin{minipage}[l]{0.41\linewidth}
\includegraphics[width=2.5in]{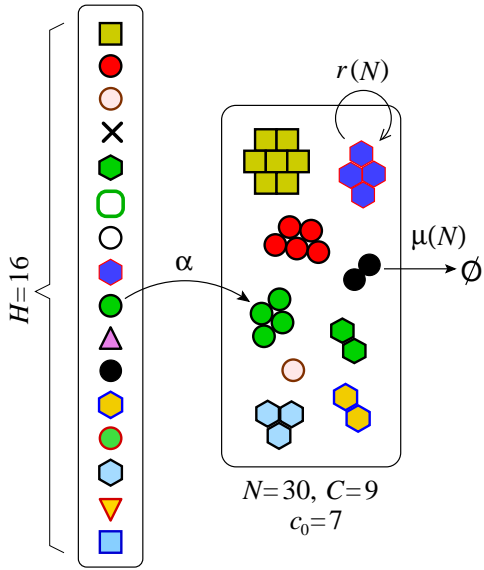}
    \end{minipage}\hfill
    \begin{minipage}[c]{0.57\linewidth}
\caption{\baselineskip=11pt A simple $H$-species
  birth-death-immigration process in which an external fixed
  ``source'' always contains $H$ individuals, each of a different
  species. This source may represent uniquely tagged stem cells, a
  ``mainland'' from which species emigrate, or the thymus that outputs
  naive T cell clone, each expressing a different T cell
  receptor. Each cell in the source buds off with rate $\alpha$ a
  daughter cell into the system but remains intact.  All individuals
  in the system can proliferate with rate $r(N)$ and dies with rate
  $\mu(N)$, where $N$ is the total population in the system. A
  specific configuration with $H=16$ and $N=30$ is depicted.  Here,
  $C=9$ represents the \textit{number of different species} that exist
  in the system. $c_{0}$ represents the number of species in the
  source that are not represented in the system.}
\label{BDI}
\end{minipage} 
\end{center}
\end{figure}
%


The simplest single-compartment mathematical structure that is common
to all the multispecies systems mentioned above is the
birth-death-immigration (BDI) processes shown in Fig.~\ref{BDI}.
%
%
The source of immigration into the system is a fixed ``source''
population of $H$ different individuals, each of a different
species. In the T cell setting, the possible number of different
receptors that can be produced by the thymus is $H> 10^{15}$
\cite{LAYDON} while in typical barcoding experiments $H\approx
10^{3}-10^{6}$ different tags can be implanted \cite{kim2014dynamics}.
After immigration into the system, the individuals can proliferate
with rate $r(N)$ and die with rate $\mu(N)$, both possibly function of
the total population $N$.  In the configuration shown in
Fig.~\ref{BDI}, the maximum number of different species is $H=16$ and
the number of individuals of each species is $n_{1} = 7, n_{2}=5,
n_{3}=n_{4}=4, n_{5}=3, n_{6}=n_{7}=n_{8}=2, n_{9}=1$. We have labeled
the species $i$ according to decreasing population. In this work, the
terms ``clones" and ``species" are interchangeable and ``clones'' will
only be used to refer to different tags in barcoding experiments and
different T-cell receptor (TCR) types.


Such high-dimensional stochastic systems are generally difficult to
study because of the ``curse of dimensionality."  The evolution of the
full probability distribution $P(\n) \equiv P(\{n_1,n_2,...,n_H\})$ is
unintuitive and computationally intractable \cite{DESSALLES}.  It also
contains more information than necessary if we consider only neutral
species and their identities are not relevant.  Describing the system
in terms of moments such as $\langle n_i \rangle$ and $\langle n_i n_j
\rangle$ reduces the model complexity and allows one to track the
dynamics of specific species \cite{goyal2015mechanisms}, but does not
directly capture the species size distribution resulting from the
relevant stochastic processes.  Another approach is to use
single-quantity metrics such as species richness,
Simpson's diversity, Shannon's diversity, or the Gini index to
describe and compare various ecological communities. Such diversity
measures can be overly simplistic and can lead to different
conclusions depending on the diversity index used.  Thus, a
description of intermediate complexity is desired.

\begin{figure}[h!]
\centering
\includegraphics[width=4.4in]{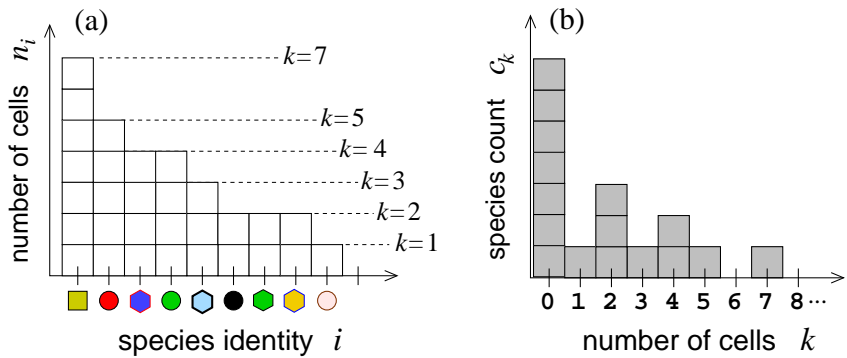}
\caption{\baselineskip=10.4pt Definition of species counts
  corresponding to the configuration in Fig.~\ref{BDI}. (a) In the
  cell-count representation $n_{i}$ is the number of cells of species
  $i$ detected in a sample. (b) ${c}_{k}$ is the number of different
  species that are represented by exactly $k$ cells in a sample. A
  given set $\{n_{i}\}$ uniquely determines the corresponding
  ${c}_{k}\equiv \sum_{i=1}^{\infty} \mathds{1}(n_{i},k)$. However,
  one cannot recover $n_{i}$ from ${c}_{k}$ since species identity
  information is lost when transforming from $n_{i}$ to ${c}_{k}$.}
  \label{fig:dos1}
\end{figure}

In ecology, a commonly used measure is the species abundance
distribution (SAD) that counts the number of different species
encountered in a community \cite{mcgill2007species}.  In the language
of clonal dynamics, it is the count of the number of species or
  ``clones'' that are each represented by $k$ individuals as depicted
in Fig.~\ref{fig:dos1} \cite{goyal2015mechanisms}:
\begin{equation}
{c}_{k} = \sum_{i=1}^{H} \mathds{1}(n_{i},k),
\label{CKDEF}
\end{equation}
Here $\mathds{1}(x,y)$ is the identity function which takes on the value 1
 when $x=y$ or 0 otherwise. 
The species-count ${c}_{k}$ represents a
one-dimensional vector of numbers indexed by $k=0,1,...$ and gives a
more comprehensive picture of how the clone/species are distributed
compared to that of a single index. By construction, the
species count $c_{k}$ also obeys the
constraints
\begin{equation}
c_{0} + \sum_{k=1}^{\infty} c_k = H,\qquad \sum_{k=0}^{\infty} k c_k = N.
\label{eq:consist}
\end{equation}
Species counts are useful in describing numbers of rare or abundant
species especially when their identities are not important.  Examples
of such as systems include genetically barcoded, virally tagged, or
TCR-decorated \cite{jin2013quantifying, zarnitsyna2013estimating,
  qi2014diversity,goyal2015mechanisms, desponds2016fluctuating}
cellular clones, microbial populations \cite{hill2003using,
  hong2006predicting}, and ecological species
\cite{10.2307/j.ctt7rj8w, mcgill2007species, guisan2005predicting}.

A universally observed feature in empirical studies across all these
fields is a ``hollow curve distribution" for $c_k$, where few highly
populous species and many low-population species arise
\cite{mcgill2007species}.  Theoretical studies have attempted to
explain these observations by proposing various physical models,
including neutral models with constant immigration, birth, and death
rates \cite{motomura1932statistical, fisher1943relation,
  kendall1948some, DESSALLES}, time-dependent birth and death
\cite{kendall1948generalized}, cell-wise and species-wise
heterogeneities \cite{desponds2016fluctuating}, and intra-species
carrying capacities \cite{volkov2005density}.

Multi-clonal/multi-species models with fluctuating total population
size commonly arise in the evolutionary biology and physics
communities \cite{parsons2008absorption, chotibut2015evolutionary,
  constable2016demographic, chotibut2017population,
  constable2017mapping}.  Most of the attention has been on computing
expected values of population counts rather than species abundance
distributions.  For example, Parsons {\it et al.}
\cite{parsons2008absorption} used a neutral and quasi-neutral
birth-death model with carrying capacity and studied the mean fixation
time of any species.

In ecology and immunology, the literature on the distribution $\langle
c_k \rangle$ is rich.  For example, Volkov et
al. \cite{volkov2005density} considered an \textit{intra-species}
carrying capacity that balances the birth and death rates of each
species. However, they did not consider a global carrying capacity
that regulates the dynamics of the population across all species.  In
this case, no interactions arise among the species and the mean-field
results are accurate.
Other theories considered competition for resources (such as T cell
proliferation competing for stimuli from self-peptides
\cite{lythe2016many,
  desponds2016fluctuating,eftimie2016mathematical}), but the resources
were modeled as evolving variables of the system and explicit
solutions could be found only in very simple cases.  None of these
previous studies have treated global interactions that correlate
populations across all clones/species, an important ingredient in
studies of interacting populations.

Stochastic simulations of a neutral BDI model that includes a simple
global carrying capacity exhibit distributions of $\langle
c_{k}\rangle$ that differ qualitatively from those of the
above-mentioned studies.  Under a high immigration rate, the classical
power law distribution of $\langle c_{k}\rangle$ (with an exponential
cutoff) remains an accurate representation of our simulated results.
However, as the immigration rate is decreased, a single
large-population species emerges.  Such single-species dominance is
not captured by classical mean-field theories.  Through further
analysis that more accurately includes the interactions between
species populations, we find that a low immigration rate induces a
phase transition to bistability in species populations.  The
mean-field approximation, which was explicitly or implicitly assumed
in previous studies, breaks down near the new local stable state where
the ensemble average of $c_k$ follows a different distribution.

%




%
%

In this paper, we introduce the simple idea of transforming the
problem of calculating the $q^{\rm th}$ moment of $c_k$ to the problem
of solving a $(q+1)$-dimensional process described by the population
vector $\{n_1,~,n_2,~...,~n_q,~N'\}$. This process can be further
approximated by a $q$-dimensional Moran model that imposes a fixed
total population.  We use this approach to accurately calculate the
$1^{\rm st}$ and $2^{\rm nd}$ moments of $c_k$ under general
functional forms of the carrying capacity.  We then exploit ideas from
energy landscapes to identify the key parameters controlling a phase
transition of the general multi-species BDI process that explains the
failure of previously used mean-field assumptions.

\section{Classical formulation and mean-field assumption}

Here, we develop the stochastic dynamics of the BDI process depicted
in Fig.~\ref{BDI}.  In the language of clonally tracked stem cell
differentiation, the probability of a stem cell carrying any specific
tag asymmetrically differentiating to produce a progenitor cell within
infinitesimal time $\dd t$ is $\alpha \dd t$. We will assume that
there is a fixed number $H$ of stem cells or ``source'' individuals.
The probability for any progenitor cell to divide into two new
identical progenitor cells (birth) within $\dd t$ is $r\dd t$
and the probability of it dying in $\dd t$ is $\mu \dd t$.

We will further assume the particle dynamics are coupled in a
species-independent way, leading to identical (but not independent)
statistics of the populations of each species.  The canonical
implementation of such a ``global'' neutral interaction is through a
birth rate $r(N)$ and/or death rate $\mu(N)$ which depend only on the
total population $N \equiv \sum_{i=1}^H n_i = \sum_{\ell}\ell
c_{\ell}$. Thus, the total population $N$ can be ``decoupled'' and
completely described by its own master equation,

\begin{equation}
\begin{array}{l}
\displaystyle {\partial  P(N,t)  \over \partial t} =  
\alpha H [P(N-1)-P(N)] \\
\:  \hspace{1.8cm} + r(N-1)(N-1)P(N-1)-r(N)NP(N) \\
\: \hspace{1.8cm} +  \mu(N+1)(N+1)P(N+1)-\mu(N)NP(N),
\end{array}
\label{eq:PN}
\end{equation}
from which moments of $N$ can be computed. The higher-dimensional
master equation obeyed by the full multispecies distribution
$P(\{n_{j}\};t)$ is explicitly given in Appendix \ref{appendix:pn}.

Let us denote the ensemble (not time) average of a quantity by
$\langle \cdot\rangle$.  Thus, $\langle n_i(t) \rangle \equiv
\sum_{\{n_j\}} n_i P(\{ n_j \}; t)$ represents the expected
population of the $i^{\rm th}$ species. By using Eqs.~(\ref{eq:PN}) and
(\ref{mastereqn:pn}), we can show that the expected subpopulation
$\langle n_{i}(t)\rangle$ and total population $\langle N(t)\rangle$
for the BDI process obeys

\begin{align}
\frac{\dd \langle n_i \rangle}{\dd t} & =  \alpha + \langle (r(N) - \mu(N))
n_i \rangle, \nonumber \\
\frac{\dd \langle N \rangle}{\dd t} & =
 \alpha H + \langle (r(N) - \mu(N)) N \rangle.
\label{DNDT}
\end{align}
In Appendix \ref{ap:ckreal}, we also explicitly derive the equation
for $\langle c_{k}(t)\rangle$,

\begin{equation}
{\dd \langle c_k \rangle \over \dd t} =  
\alpha (\langle c_{k-1} \rangle - \langle c_k \rangle) 
+ \langle r(N)\left[(k-1) c_{k-1}- k c_k\right]\rangle
+  \langle \mu(N)\left[(k+1) c_{k+1} - k c_k\right] \rangle,
\label{Eq:truemean}
\end{equation}
from the master equation for $P(c_{0},c_{1},c_{2},\ldots; t)$.  This
evolution equation indicates that immigration (at rate $\alpha$) of an
individual from a species with population $n_i=k$ increases its size
by 1, thereby decreasing $c_{k}$ by 1 but increasing the number of
species with population $k+1$, $c_{k+1}$, by 1.  Cellular birth and
death have similar effects, but their corresponding rates are
proportional to the species population $k$ (the number of
individuals/cells in the species).
%
In the rest of this paper we will be interested in evaluating the
steady-state values of $\langle c_k \rangle$.

\subsection{Constant rates} 

\noindent In the simplest scenario of constant birth and death rates, one can
write \cite{goyal2015mechanisms}

\begin{equation}
{\dd \langle c_k \rangle \over \dd t} =  
\alpha (\langle c_{k-1} \rangle - \langle c_k \rangle) 
+ r [(k-1) \langle{c_{k-1}}\rangle - k\langle{c_k}\rangle]
+  \mu [(k+1) \langle c_{k+1} \rangle - k \langle c_k \rangle].
\label{eq:ck-const}
\end{equation}
If $r<\mu$, a stable steady state can be found:


\begin{equation}
\langle{c_{k\geq 1}^*}\rangle = \frac{\alpha H}{r k!}\frac{(\frac{r}{\mu})^k (1-\frac{r}{\mu})^{\alpha/r}}{\frac{\alpha}{r}+k} 
\prod_{\l=1}^k \left( \frac{\alpha}{r} +\l \right), \quad\,
\langle {c}_0^* \rangle = H - \sum_{k=1}^\infty \langle{c_k^*}\rangle = H \left( 1-\frac{r}{\mu} \right)^{\alpha/r}.
\label{Eq:goyal}
\end{equation} 
In the $\alpha/r \to 0^{+}$ limit, the species counts monotonically
decay as

\begin{equation}
\langle c_{k\geq 1}^{*}\rangle \approx H\left({\alpha \over
  r}\right)\left(1-{r \over \mu}\right)^{\alpha/r}\left({r\over
  \mu}\right)^{k}{1\over k}.
\label{CK0}
\end{equation}

\subsection{Carrying capacity and mean-field approximation}
Now, assume that $r(N)$ decreases with $N$ and/or $\mu(N)$ increases
with $N$, and that $\lim_{N\to \infty} r(N)/\mu(N) < 1$.  These
conditions on $r(N)$ and $\mu(N)$ guarantee that $n_i$ and $c_k$ are
bounded even if $r(N) > \mu(N)$ for some finite $N$. Terms of the form
$\langle {r(N) c_{k}} \rangle$ in Eq.~(\ref{Eq:truemean}) cannot be
approximated by factoring because $r(N)$ depends on $c_k$ through the
stochastic variable $N\equiv \sum_\l \l c_\l$ defined in
Eq.~(\ref{eq:consist}).  Nonetheless, to make headway, a mean-field
method is often invoked to simplify Eq.~(\ref{DNDT}) and
Eq.~(\ref{Eq:truemean}). Upon fully factorizing interaction terms such
as $\langle r(N) c_{k}\rangle \approx r(\langle N\rangle)\langle
c_{k}\rangle$ and $\langle r(N) N\rangle \approx r(\langle
N\rangle)\langle N \rangle$, we can approximate Eqs.~(\ref{DNDT}) and
(\ref{Eq:truemean}) as
\begin{align}
\frac{\dd \langle N \rangle}{\dd t} 
\approx &~ \alpha H + \langle (r(\langle N\rangle) - \mu(\langle N\rangle)) 
\langle N \rangle \equiv f(\langle N\rangle), \label{DNDT_MFT} \\
{\dd \langle c_k \rangle \over \dd t}  \approx &~ \alpha (\langle c_{k-1}
\rangle - \langle c_k \rangle) + r(\langle N\rangle)\left[(k-1)
  \langle c_{k-1}\rangle - k \langle c_k \rangle\right] \nonumber \\
\:  & \quad + \mu(\langle
N\rangle) \left[(k+1)\langle c_{k+1} \rangle - k \langle c_k
  \rangle\right].
\label{Eq:mf-dynamics}
\end{align}

By first solving Eq.~(\ref{DNDT_MFT}) we can input $\langle
N(t)\rangle$ into Eq.~(\ref{Eq:mf-dynamics}) and explicitly solve for
$\langle c_{k}(t)\rangle$.  The steady-state solution to $\langle
N\rangle$, $\langle N^{*}\rangle$, is defined in
Eq.~(\ref{DNDT_MFT}) by $f(\langle N^{*}\rangle) = 0$ and the
requirement that $\langle N^*\rangle >0$ requires $[d f(\langle
  N\rangle)/d\langle N\rangle]_{\langle N^{*}\rangle} \equiv
f'(\langle N^{*}\rangle) \equiv r'(\langle N^{*}\rangle)- \mu'(\langle
N^{*}\rangle) < 0$.  The steady state values of $\langle
c_k\rangle$ can be reached only after the steady state of $\langle
N\rangle$ is reached and $r(\langle N\rangle)$ and $\mu(\langle
N\rangle)$ approach constant values.

%

We show in Appendix \ref{Ap:multi} that this deterministic description
breaks down after an exponentially long time when the immigration rate
$\alpha$ is sufficiently small.  The reason is that for $\alpha = 0$,
$N=0$ becomes an absorbing boundary in the full stochastic model.
Thus, when $\alpha = 0$, the $\langle N^*\rangle$ we find from
$f(\langle N^{*}\rangle) = 0$ is actually a quasi-steady state (QSS)
even though Eq.~(\ref{DNDT_MFT}) indicates a stable deterministic
equilibrium $\langle N^{*}\rangle$ for physically reasonable functions
$r(\langle N\rangle)$ and $\mu(\langle N\rangle)$.

Focusing on evaluating the QSS value of $\langle c_k \rangle$,
$\langle c_{k}^{*}\rangle$, before the final extinction that occurs
over exponentially long times, we denote $r(\langle N^*\rangle) \equiv
r^*$ and $\mu(\langle N^*\rangle) \equiv \mu^*$ as the rates of birth
and death at QSS. The QSS solution $\langle c_k^*\rangle$ can be
written in the same form as Eq.~(\ref{Eq:goyal}),

\begin{equation}
\langle c_{k\geq 1}^*\rangle  = \frac{\alpha H}{r^* k!}
\frac{(\frac{r^*}{\mu^*})^k (1-\frac{r^*}{\mu^*})^{\alpha/r^*}}{\frac{\alpha}{r^*}+k} 
\prod_{\l=1}^k \left( \frac{\alpha}{r^*} +\l \right)
,\quad
\langle c_0^*\rangle  = H - \sum_{k=1}^\infty \langle c_k^*\rangle 
= H \left( 1-\frac{r^*}{\mu^*} \right)^{\alpha/r^*}\!\!\!\!.
\label{Eq:mf}
\end{equation} 
Here, $\langle c_{k}^{*}\rangle$ corresponds to the mean QSS
species-count under the mean-field approximation which we expect to be
different from the exact solution. In the $\alpha/\mu^{*},
\alpha/r^{*}\to 0^{+}$ limit, the expected species count $\langle
c_{k\geq 1}^{*}\rangle$, as in Eq.~(\ref{CK0}), is monotonic in $k$:

\begin{equation}
\langle c_{k\geq 1}^{*}\rangle \approx H\left({\alpha \over
  r^{*}}\right)\left(1-{r^{*} \over \mu^{*}}\right)^{\alpha/r^{*}}\left({r^{*}\over
  \mu^{*}}\right)^{k}{1\over k}.
\label{CK00}
\end{equation} 
However, under regulation, $r^{*}/\mu^{*} \approx 1 - {\cal
  O}(\alpha/\mu^{*})$ resulting in a long-tail $k$-dependence of
$\langle c_{k}^{*}\rangle$. Although the amplitude of $\langle
c_{k\geq 1}^{*}\rangle$ is proportional to $\alpha/r^{*}$, it is
constructed to obey the mean total population constraint $\langle
N^{*}\rangle = \sum_{k=1}^{\infty}k \langle c_{k}^{*}\rangle$ which is
reflected in the long-tail property of the mean-field approximation to
$\langle c_{k}^{*}\rangle$.

\subsection{Failure of the mean-field approximation to $\langle c_k^*\rangle$ 
in the slow immigration regime}

To concretely investigate the errors incurred under a mean-field
assumption, we first focus explicitly on a logistic growth law for the
total population defined by

\begin{equation}
r(N) = p \left( 1-\frac{N}{K} \right),\qquad  
\mu(N) = \mu,
\label{eq:logi-setup}
\end{equation}
where $p$ is the maximal birth rate and $K$ is the carrying capacity
parameter. The mean-field solution
for the total population is

\begin{align}
\langle N^*\rangle &  = \frac{K}{2}\left(1-{\mu \over p}\right)
\left[1+\sqrt{1+\frac{4\alpha H p}{(p-\mu)^{2}K}}\right] \nonumber \\
\: & = K\left(1-{\mu \over p}\right) + \frac{\alpha H p}{p-\mu} + {\cal O}(1/K).
\label{eq:ssLogi}
\end{align}
In many examples, such as progenitor cells, $K \gg 1$ and $\langle
N^*\rangle \sim K$ is large except when $\mu$ approaches or exceeds
$p$.

\begin{figure}[h!]
\centering
\includegraphics[width=7in]{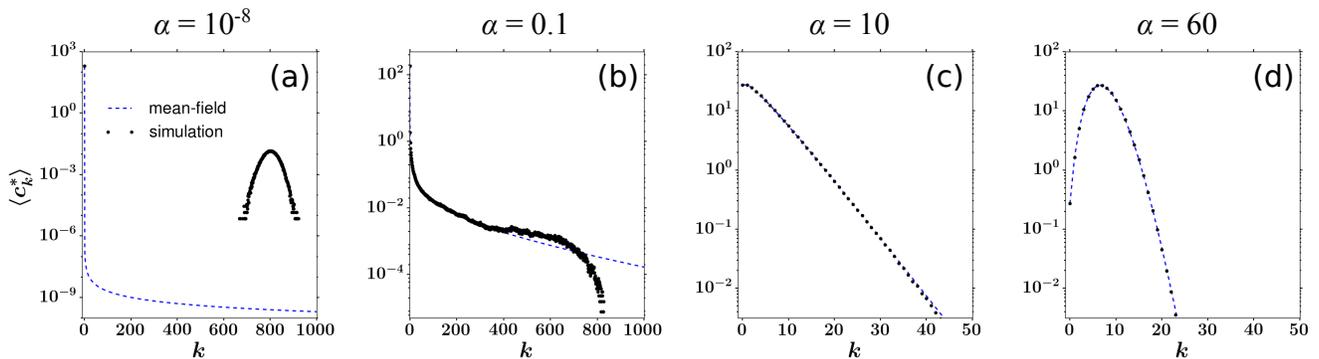}
\caption{Comparison of steady-state species-count distributions from
  simulations (black dots) to those computed from the mean-field
  approximation (dashed blue curves) using the logistic growth model
  of Eq.~(\ref{eq:logi-setup}) and (a) $\alpha = 10^{-8}$, (b) $\alpha
  = 0.1$, (c) $\alpha = 10$, and (d) $\alpha = 60$.  Other parameters
  used are $\mu=10,~p=20,~K=1600,~H=200$. The resulting $N^*$ are
  $800,~ 840,~ 4800$, and $24800$, respectively.  The mean-field
  approximation $\langle c_{k}^{*}\rangle$ breaks down for small
  $\alpha$ completely missing the peak at $k \approx \langle
  N^{*}\rangle$ in (a). Also, note the log scale and the absence of
  simulations that capture the rare configurations in (a) where $k
  \neq \langle N^{*}\rangle$.}
\label{fig:goyal}
\end{figure}

We are now in a position to use $\langle N^*\rangle$ to determine
$r^*$ and evaluate the mean-field approximation for $\langle
c_k^*\rangle$ (Eq.~(\ref{Eq:mf})).  In Fig.~\ref{fig:goyal} we compare
numerically evaluated mean-field solutions of $\langle c_k^*\rangle$
with Monte-Carlo simulations of the underlying BDI process for various
values of $\alpha$.  

For small $\alpha$, such as $10^{-8}$ used to generate
Fig.~\ref{fig:goyal}(a), Eq.~(\ref{Eq:mf}) fails to capture the peak
arising in $\langle c_k^* \rangle$ at $k\approx\langle N^*\rangle$. In
the singular limit $\alpha\rightarrow 0$, the mean-field solution
$\langle c_{k \geq 1}^*\rangle \rightarrow 0$ and $\langle
c_0^*\rangle \rightarrow H$ but nonetheless, by construction,
satisfies $\sum_{k=1}^{\infty} k \langle c_k^* \rangle \rightarrow
\langle N^{*}\rangle$.  However, in the simulated
$\langle {c}_{k}^{*} \rangle$, the small peak at large size $k\approx
\langle N^{*}\rangle$ signals that a single species has come to
dominate the total population. The number of species not in the system
is thus $\langle {c}_0^* \rangle \approx H-1$. One species, typically
the first to have immigrated, has taken over the system squeezing out
all others that try to immigrate when the immigration rate $\alpha$ is
small. This peak in $\langle c_{k}^{*}\rangle$ near $k \approx \langle
N^{*}\rangle$ is completely missed by the mean-field approximation.
The mean-field approximation also inaccurately captures the rapid
decay in $\langle c_{k}^{*}\rangle$ for $k > \langle N^{*}\rangle$ due
to exhaustion of the population in the single size-$k$ species.

%
%
When $\alpha$ is still relatively small as in Fig.~\ref{fig:goyal}(b),
the simulated $\langle c_{k}^{*}\rangle$ is dominated by many
low-population species, with a slow decay with size $k$ followed by a
faster decay in $k$, again due to mass depletion.  At this modest
immigration rate, high-population species do not have the opportunity
to establish and more intermediate-sized species arise at the expense
of very high-population species but the simulated result $\langle
c_{k}^{*}\rangle$ remains monotonic.  Nevertheless, the mean-field
approximation of Eq.~(\ref{Eq:mf}) still fails to capture the fast
decay of $\langle c_{k}^{*}\rangle$ for large $k$.

For even larger $\alpha$, the preferred total population increases.
Since the total number of species remains capped at $H$, the mean
number of individuals/cells per species increases. The distribution
$P(n_{1})$ peaks at higher values of $n_1$ thereby forming a peak in
$\langle c_{k}^{*}\rangle$ at size $k \ll \langle N^{*}\rangle$. The
larger-$\alpha$ cases shown in Fig.~\ref{fig:goyal}(c-d) are
accurately described by the mean-field approximation of $\langle
c_{k}^{*}\rangle$ for all values of $k$.

\section{Proposed Model for $\langle \c^q \rangle$}
\label{sec:methods}
The challenge in solving Eq.~(\ref{Eq:truemean}) lies in the
nonseparable terms $\langle r(N) c_k \rangle$.  Even in the simple
case of logistic growth where $r(N)$ is linear, the $\langle r(N) c_k
\rangle$ terms include second-moments $\langle c_k c_{\ell}\rangle$,
which usually cannot be approximated by $\langle c_k \rangle \langle
c_{\ell} \rangle$.  If one attempts to solve Eq.~(\ref{Eq:truemean})
for the time-dependent or steady-state solution $\langle c_k^*
\rangle$, one encounters the so-called ``moment closure" problem,
where the solution of the $1^{\rm st}$ moment $\langle c_k \rangle$
depends on $2^{\rm nd}$ moments $\langle c_k c_\l \rangle$, which in
turn depends on $3^{\rm rd}$ moments, and so on
\cite{COLIN2009}. There is usually no closed-form solution or easy
approximation to such problems.  In the rest of this section, we
develop an alternative approach.

\subsection{Transformation of the problem}
A complete description would be an $H$-dimensional model for the
distribution $P(\{n_{1}, n_{2}, \ldots, n_{H}\};t)$. However, by using
the definition of $c_{k}$ in Eq.~(\ref{CKDEF}), assuming the initial
populations of all species are identical $n_{1}(0)=n_{2}(0) = \ldots =
n_{H}(0)$ and indistinguishability among species, one can easily show
that (Appendix \ref{Ap:Moments})

\begin{equation}
\langle {c}_{k}(t) \rangle = HP(n_1=k;t),
\label{eq:ck-true}
\end{equation}
where $P(n_{1};t) =
\sum_{n_{2}=0}^{\infty}\cdots\sum_{n_{H}=0}^{\infty} P(\{n_{1}, n_{2},
\ldots, n_{H}\};t)$ is the single-dimensional marginal distribution.
Here, the singling out of species 1 is arbitrary.  The assumption of
identical initial species populations is not needed in the long time
QSS limit as long as different initial distributions $c_{k}(0)$
converge to a unique $\langle c_k^* \rangle$.  Thus, at QSS $\langle
{c}_{k}^{*} \rangle = HP(n_1=k;t\to\infty)$ always holds.
Intuitively, the expected fraction of all species that have size $k$
is the probability that any one species is of size $k$.
%
%

%

We can write the master equation for the BDI process of a single species
with population $n_1$ as
%

\begin{equation}
\begin{array}{l}
\displaystyle {\partial  P(n_1;t)  \over \partial t} =  
\alpha [ P(n_1 - 1)  -  P(n_1) ] + r(N)
\left[(n_1 - 1) P(n_{1} - 1) -  n_{1} P(n_1)\right] \\
\: \hspace{1in} +  \mu(N)\left[(n_{1}+1)  P(n_1+1)  -  n_1 P(n_1)\right]
\end{array}
\label{eq:Pn}
\end{equation}
where $N(t)$ represents one trajectory of the
random process $N=n_{1}+n_{2}+\ldots + n_{H}$ which we might
approximate using the deterministic solution to Eq.~(\ref{DNDT}).
Equation (\ref{eq:Pn}) has the exact same form as the 
right-hand side of Eq.~(\ref{Eq:mf-dynamics}) for $\langle c_k \rangle$.
However, in the presence of other species or clones, it is immediately
clear that Eq.~(\ref{eq:Pn}) is not a complete description for $n_1$
since the variable $N$ depends on the population of all
species. Species ``independence" breaks down through the $r(N)$ and
$\mu(N)$ terms.  All species compete with each other for the limited
sources in the environment through their shared and regulated birth
and death rates.


Eq.~(\ref{eq:ck-true}) remains exact (Appendix \ref{Ap:Moments}) since
the population dynamics are neutral and all species start with the
same initial size.  One still needs to solve any \textit{individual}
species' marginal probability distribution $P(n_1)$ given that all
species, including itself can affect it.  Formally, this corresponds
to first solving the full distribution $P(\{n_1,~n_2,~...,~n_H\})$
before summing over all other populations $\{n_2,~...,~n_H\}$.  Since
we are not concerned about the detailed configurations of
$\{n_2,~...,~n_H\}$, but rather their combined effects on $n_1$.
Therefore, we can lump species 2 through $H$ into an effective ``bath"
species whose size is $N' = n_2+...+n_H$.  This effective
species has a birth rate $N' r(n_1+N')$, a
death rate $N'\mu(n_1+N')$, and an immigration rate
$(H-1)\alpha$.  Eq.~(\ref{eq:Pn}) is now coupled to the master
equation

\begin{align}
  {\partial P(N';t)  \over \partial t} =  &\,
  \alpha(H-1) [ P(N' - 1)  -  P(N') ] + r(N)\left[
(N' - 1)P(N' - 1) - N' P(N')\right] \nonumber \\
\:&  \quad +  \mu(N)\left[(N'+1)  P(N'+1)  -  N' P(N')\right].
\label{eq:PnOther}
\end{align}
One usually combines Eqs.~(\ref{eq:Pn}) and (\ref{eq:PnOther})
together into a 2D master equation

\begin{align}
\frac{\partial P(n_1,N';t)}{\partial t} = &
\, \alpha [ P(n_1 - 1, N')  -  P(n_1, N') ] \nonumber \\
\: & + \alpha(H-1) [ P(n_1, N' - 1)  -  P(n_1, N') ] \nonumber \\
\: & + r(N-1)\left[ (n_1 - 1) P(n_1 - 1, N') + (N' - 1) P(n_1, N' - 1) \right] \nonumber \\
\:& +  \mu(N+1)\left[(n_1+1)  P(n_1+1, N') + (N'+1)  P(n_1, N'+1) \right] \nonumber \\
\: & - [r(N) + \mu(N)] \left[ n_1 P(n_1, N') + N' P(n_1, N')\right].
\label{eq:Pn2D}
\end{align}
The 2D problem can be approximated by a 1D problem when $n_1 \ll N'$
and $r(N) \approx r(N')$.  The birth rate is approximately regulated
by the ``bath'' population $N'$ which leads to a decoupling
from $n_1$.  Similarly, when $n_1 \gg N'$, $r(N) \approx r(n_{1})$ and
the birth rate is approximately independent of $N'$.  In either limit,
the problem is approximately one-dimensional and can be modeled using
a 1D master equation for $P(n_1)$ [Eq.~(\ref{eq:Pn})] or $P(N')$
[Eq.~(\ref{eq:PN})] correspondingly.  However, when $n_1$ and $N'$ are
comparable in size, one needs to evaluate the full 2D distribution
$P(n_1,N')$ and marginalise over $N'$ to obtain $P(n_1) =
\sum_{N'=0}^{\infty} P(n_1,N')$ and
\begin{equation}
\langle c_k(t) \rangle = HP(n_1=k;t) = H\sum_{N'=0}^{\infty} P(n_1=k,N';t).
\end{equation}

This approach can be extended to higher dimensions to determine higher
moments of $c_k(t)$, which are important for characterizing the
variability of species size distributions.  Covariances ${\rm
  cov}(c_k,c_\l)\equiv \langle {c}_{k} {c}_{\l} \rangle - \langle
{c}_{k} \rangle \langle {c}_{\l} \rangle$, in particular, will reveal
the differences between the solutions to the mean-field model
[Eq.~(\ref{Eq:mf})] and the exact model [Eq.~(\ref{Eq:truemean})]. In
Appendix \ref{Ap:Moments}, we derive relationships between higher
moments of $c_k$ and the cell count distributions $P(n_1,n_2,...)$.
Specifically, for the second moments,
\begin{equation}
\langle {c}_{k}(t) {c}_{\ell}(t)\rangle = 
H(H-1) P(n_{1} = k, n_{2} = \ell; t) + \mathds{1}(k,\ell) H P(n_{1}=k; t).
\label{eq:ckcl}
\end{equation}


\subsection{Approximating $P(\{n_1, n_2, ..., n_q, N'\})$ 
by a $q$-dimensional Moran model}

We now try to find a solution to $P(n_1, N')$.  Since the 2D master
equation does not usually have analytic solutions, we will show how to
approximate $P(n_1,N')$ by a 1D two-species Moran model
\cite{parsons2008absorption, chotibut2015evolutionary,
  constable2016demographic, chotibut2017population,
  constable2017mapping} with $n_{1}$ individuals of species 1 and $N'$
individuals of species 2 (which, for this case, is the sum of the
populations of species 2 through $H$ in the original multispecies
model).  The 1D Moran model imposes $n_1 + N' \equiv N$, the total
population size, to be a fixed value.

We first fix the value of $N$ to be the quasisteady-state value of the
original unconstrained BDI process $N \to N^{*} \coloneqq \langle
N^{*}\rangle$, at which the condition $\alpha H + r(N^*) N^* =
\mu(N^*) N^*$ is satisfied.  For example, under a logistic birth law
   [Eq.~(\ref{eq:logi-setup})], the mean-field approximation
   Eq.~(\ref{eq:ssLogi}) yields an accurate value of $N^{*}$.  At this
   value of $N^{*}$, the growth and death rates take on specific
   values defined by $r^*\coloneqq r(N^*),~\mu(N^*)\coloneqq \mu^*$.
   In fact, to absolutely fix $N^*$ the stochastic dynamics are driven
   by completely coupled birth and death events.  During each event,
   one individual is randomly chosen to die and immediately replaced
   by a new one.  This tethering of birth and death ensures that the
   total population $N^*$ is fixed.  The total rate of a tethered
   birth-death event is $\frac{1}{2}(\alpha H + r^* N^* + \mu^* N^*) =
   \mu^*N^*$, where the factor $1/2$ factors in the fact that two
   birth-death events occur simultaneously during one tethered event
   so on average the arrival rate of events has to be halved. Thus,
   $\mu^*N^*$ is the intrinsic rate of evolution in the Moran model.
   The master equation for the probability distribution $P_{\rm
     M}(n_{1};t\vert N^{*})$ of the fixed-$N^*$ two-species Moran
   model can be expressed as
%

\begin{align} 
\frac{\partial P_{\rm M}(n_{1};t \vert N^*)}{\partial t}  = & \omega_{12}(n_1-1\vert N^*) 
P_{\rm M}(n_{1}-1\vert N^*)
+ \omega_{21}(n_1+1\vert N^*) P_{\rm M}(n_1+1\vert N^*) \nonumber \\
\: & - \left[\omega_{12}(n_{1}\vert N^*)+ 
\omega_{21}(n_{1}\vert N^*)\right]P_{\rm M}(n_{1}\vert N^*),
\label{eq:Moran-master}
\end{align}
where the functions $\omega_{ji}(n\vert N^*)$ denote the rate that a
species-$i$ individual is replaced by a species-$j$ individual in a
Moran process of fixed total population $N^*$
\begin{align}
\omega_{12}(n\vert N^*) & =  n\left(1 - \frac{n}{N^*} \right)r^{*} +
\left(1 - \frac{n}{N^*}\right)\alpha \nonumber \\
\: & = \mu^* N^* \left[(1-m^{*}){n\over N^{*}}\left(1-{n \over N^{*}}\right) 
+ m^{*}Q_1\left(1 -{n\over N^{*}}\right) \right],
\nonumber \\
\omega_{21}(n\vert N^*) & =  n\left(1 - \frac{n}{N^*} \right)r^{*} + 
(H-1)\left(\frac{n}{N^*}\right)\alpha \nonumber \\
\: & = \mu^* N^* \left[(1-m^{*}){n \over N^{*}}\left(1-{n\over N^{*}}\right) 
+ m^{*}(1-Q_1)\left({n\over N^{*}}\right) \right],
\label{eq:1d_fb}
\end{align}
where we have further defined 

\begin{equation}
m^{*}\equiv \frac{\alpha H}{\mu^* N^*}, \quad Q_1 = \frac{1}{H}.
\label{mQ}
\end{equation}
Here, $m^{*}$ represents the relative total immigration rate and $Q_1$
is the fixed fraction of species 1 amongst those in the
immigration source.
%

In these dynamics, it is clear that the probability of choosing an
individual for removal/death from species 1 and species 2 (the bath
species) are $n_{1}/N^{*}$ and $1-n_{1}/N^{*}$, respectively. The
newly created (from birth) individual has probability $n_{1}/N^{*}$ to
be of species 1 and $1-n/N^{*}$ to be of species 2, calculated from
the state of the model prior to death.
Thus, after one event, the population of species 1 may increase by 1
(if a species-2 individual is chosen to die, and a species-1
individual is chosen to be born) or decrease by 1 (if a species-1
individual is chosen to die, and a species-2 individual is chosen to
be born).  The total rate of population change of any one species
includes the per-cell immigration rate $\alpha$, which is equal to the
per-species immigration rate since the cells initiating immigration
are unique (see Fig.~\ref{BDI}). The total immigration into the
``bath'' species (species 2) is thus $(H-1)\alpha$.

%
%

To solve Eq.~(\ref{eq:Moran-master}) in steady state, we use
Eqs.~(\ref{eq:1d_fb}) and invoke the detailed balance condition
$\omega_{12}(n_1-1\vert N^*) P_{\rm M}^*(n_1-1\vert N^*) =
\omega_{21}(n_1\vert N^*) P_{\rm M}^{*}(n_1\vert N^*)$ to obtain

\begin{equation}
P^*_{\rm M}(n_1\vert N^*) = P^*_{\rm M}(0\vert N^*){\omega_{12}(0\vert N^{*})\over
\omega_{21}(n_{1}\vert N^{*})}\prod^{n_1-1}_{\l=1} 
\frac{\omega_{12}(\l\vert N^*)}{\omega_{21}(\l\vert N^*)},\quad 
P^*_{\rm M}(0\vert N^*) = \left[ \sum^{N}_{n_1=0} \prod^{n_1}_{\l=1} 
\frac{\omega_{12}(\l-1\vert N^*)}{\omega_{21}(\l\vert N^*)} \right]^{-1}. 
\qquad
\label{eq:moran1d}
\end{equation}

%


%

For general $q$-dimensional ($q\geq 2$) Moran models that involve
$(q+1\geq 3)$ subpopulations, closed-form solutions are difficult to
obtain. However, we can approximate these models using a diffusion
approximation that treats the species fractions $x_i = n_{i}/N^*$
($1\leq i \leq q$) as continuous variables. After Taylor-expanding
$q-$dimensional discrete master equations and assuming $m^{*} \equiv
{\alpha H\over \mu^*N^*} \ll 1$, a simple $q-$dimensional
Fokker-Planck equation can be derived
\cite{kimura1964,blythe2007stochastic}

\begin{equation}
\frac{\partial P_{\rm M}(\x\vert N^*)}{\mu^{*}\partial t} +  
\sum_{i=1}^{q}  \frac{\partial \left[A_i(\x) P_{\rm M}(\x\vert N^*)\right]}{\partial x_i}
= \frac{1}{N^*} \sum_{i=1}^{q} \sum_{j=1}^{q} 
\frac{\partial^2\left[B_{ij}(\x)P_{\rm M}(\x\vert N^*)\right]}{\partial x_i \partial x_j}
\label{eq:diffu}
\end{equation}
where 
\begin{eqnarray}
&&
A_i(\x) = \sum_{j=1}^{q} m^{*}(Q_i - x_i), ~~  
B_{ii}(\x) = x_i (1-x_i), ~~ B_{ij}(\x)=-x_ix_j  ~ (i\neq j). 
\end{eqnarray}
For example, when $q=2$ (three species), we
have $Q_1 = Q_2 = \frac{1}{H},~ Q_3 = \frac{H-2}{H}$.  We explicitly
show the derivations for the 1D and 2D Fokker-Planck equations in
Appendix \ref{ap:diffusion}.
The exact steady-state solution of the general
$q$-dimensional diffusion model is known and follows the Dirichlet
distribution \cite{baxter2007exact}
\begin{eqnarray}
P_{\rm M}^*({\bf n}\vert N^*) = \Gamma(N^* m^{*}) \prod_{i=1}^{q+1} \frac{(n_i/N^{*})^{N^* m^{*}Q_i - 1}}
{\Gamma(N^* m^{*}Q_i)}.
\label{Eq:P-moran-conti-solu}
\end{eqnarray}

\subsection{Relaxing the fixed-population constraint of the Moran model}
\label{sec:conv1d}
While the Moran model can be used to approximate $P^{*}(n_1, N')$, it
includes an additional hard constraint $n_1+N'=N^*$ that is not
imposed in the original BDI model.  In fact, $n_1$ itself can
fluctuate above $N^*$.  To relax this fixed-population constraint and
find an improved approximation to the reduced QSS distribution
$P^*(\{n_1,n_2,...,n_{q}\})$, we simply allow the system size of the
Moran process to vary and weight each QSS Moran process by the
steady-state probability distribution

\begin{equation}
P^*(N) = \displaystyle \frac{\prod^{N}_{j=1} \frac{r(j-1) + \alpha H}{\mu(j)}}
{\sum^{\infty}_{m=0} \prod^{m}_{\ell=1} 
\frac{r(\ell-1) + \alpha H}{\mu(\ell)}},
\label{Eq:PNsolu}
\end{equation}
which is readily obtained from solving Eq.~(\ref{eq:PN}), the master
equation for the total population of the BDI process.
%
%
%
We thus use a whole family of Moran models, each at a different value
of $N$, weighted by $P^*(N)$ to approximate the QSS probability

\begin{eqnarray}
P^*(\{n_1,n_2,...,n_{q}\}) = \sum_{N=1}^\infty P^{*}_{\rm M}
\left(\{n_1,n_2,...n_{q+1}\}\vert N \right) P^*(N). 
\label{Eq:p-convl} 
\end{eqnarray}
%
%
%

Different values of the system size will yield different values of the
rates $\omega_{ji}(n\vert N)$ according to Eq.~(\ref{eq:1d_fb}). In
1D, according to Eq.~(\ref{eq:1d_fb}), the ratio
$\frac{\omega_{12}(\l\vert N)}{\omega_{21}(\l\vert N)}$ varies with
$N$ according to

%

\begin{equation}
\frac{\omega_{12}(\l\vert N)}{\omega_{21}(\l\vert N)} = 
\frac{(1-m^{*}){\ell \over N}\left(1-{\ell\over N}\right) + 
m^{*} Q_1 \left(1-{\ell \over N}\right)}
{(1-m^{*}){\ell \over N}\left(1-{\ell\over N}\right) + m^{*}(1-Q_1)
\left({\ell\over N}\right)},
%
\label{eq:tmpOmegaRatio}
\end{equation}
where we have kept the intrinsic rates $r^{*}$ and $\mu^{*}$ and the
relative immigration rate $m^{*}$ \textit{fixed}. The only terms in
Eq.~(\ref{eq:tmpOmegaRatio}) that vary with $N$ are the relative
populations $\ell/N$ and $1-\ell/N$ reflecting only the changes
associated with changes in system size.  By keeping the $r^*, \mu^*$,
and $m^{*}$ fixed, we preserve the relative tethered rates of birth,
death, and immigration that define the original BDI process.

\section{Results}
\label{sec:results}
\subsection{$\langle c_{k} \rangle$ and $\langle c_{k}c_{\ell} \rangle$ under logistic growth}
\label{sec:res1}
\begin{figure}[h!]
\centering
\includegraphics[width=7in]{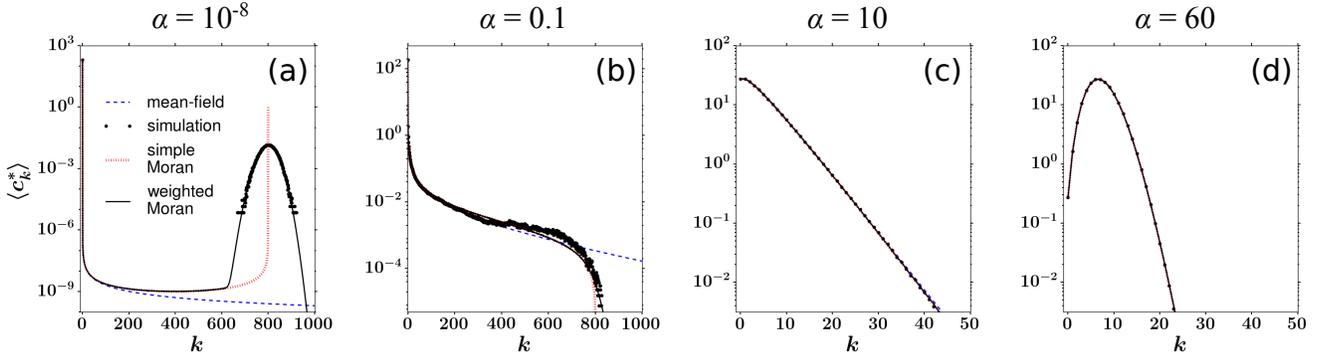}
\caption{Simulated (black dots), mean-field (blue dashed), simple
  Moran (red hashes), and weighted Moran (black solid) approximations
  of $\langle c_{k}^{*}\rangle$ using logistic growth laws and the
  parameters $\mu=10,~p=20,~K=1600,~H=200$.  Immigration rates used
  were (a) $\alpha = 10^{-8}$, (b) $\alpha = 0.1$, (c) $\alpha = 10$,
  and (d) $\alpha = 60$, as in Fig.~\ref{fig:goyal}.  In (a) we show
  the prediction from a single Moran model of fixed size $N^{*}$.  The
  weighted QSS Moran model approach (solid black curves) yields a very accurate
  approximation to the simulated values of $\langle c_{k}^{*}\rangle$
  for all values of $\alpha$, including small $\alpha$ as shown in
  (a) and (b).}
\label{fig:moran1d}
\end{figure}

In Fig.~\ref{fig:moran1d}, we plot results from Monte-Carlo
simulations, mean-field solutions to Eq.~(\ref{Eq:mf}), numerical
solutions of the simple Moran model Eq.~(\ref{eq:moran1d}), and the
weighted Moran model defined by Eqs.~(\ref{Eq:p-convl}),
(\ref{eq:moran1d}), and (\ref{eq:tmpOmegaRatio}).  As shown by
Fig.~\ref{fig:moran1d}(a), the simple Moran model has a sharp peak at
$N^*$ arising from the fixed-population constraint.  The improved
weighted solution yields accurate expected QSS
species count distributions $\langle
c_{k}^{*}\rangle$ for all values of $\alpha$, capturing the the peak
for extremely small $\alpha$ as well as the fast decay at large $k$.

%

\begin{figure}[h!]
\centering
\includegraphics[scale=0.66]{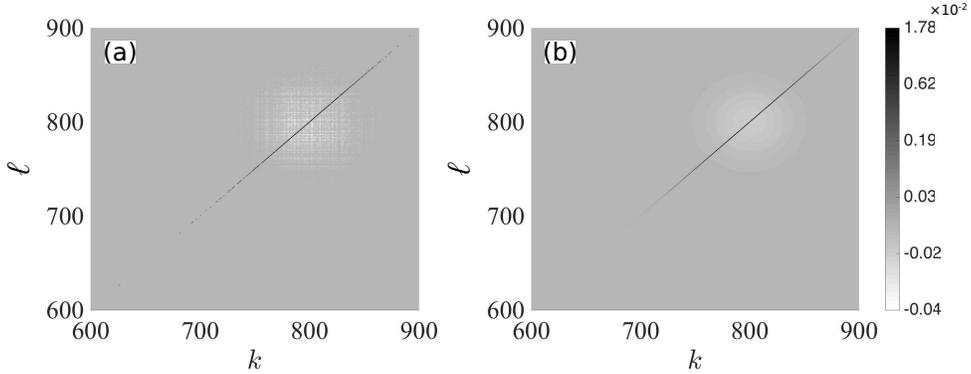}
\caption{${\rm cov}(c_k, c_\l)$ from simulations (a) and from our
  calculations (b). Parameters are $\alpha = 10^{-8},~
  \mu=10,~p=20,~K=1600,~H=200$.  Only the interesting ranges of $k$
  and $\l$ close to $N^*\approx 800$ are shown.  The pattern shows
  that large species counts are positively self-correlated (black
  line) but are negatively correlated with neighboring counts (white
  dots). The grey background shows no correlation between species
  counts of significantly different population levels.  Greyscale
  values are shown on an exponential scale.}
\label{fig:moran2d}
\end{figure}

To calculate the covariance between $c_k^{*}$ and $c_{\l}^{*}$ at QSS,
we use the 2D ($q=2$) ``continuum'' solution given in
Eq.~(\ref{Eq:P-moran-conti-solu}) in the weighting in
Eq.~(\ref{Eq:p-convl}) in order to numerically compute
Eq.~(\ref{eq:ckcl}). The covariances ${\rm cov}(c_{k}^{*}, c_{\l}^{*})
\equiv \langle c_{k}^{*} c_{\ell}^{*} \rangle - \langle c_{k}^{*}
\rangle \langle c_{\l}^{*} \rangle$ with $\alpha = 10^{-8}$, both from
Monte-Carlo simulations and from our weighted Moran model
approximation, are plotted in Fig.~\ref{fig:moran2d}. The results
provide insight on how the true dynamics for $\langle c_{k}^{*}
\rangle$ in Eq.~(\ref{Eq:truemean}) differs from that of the
mean-field description in Eq.~(\ref{Eq:mf}).

The large values (black line) along the diagonal $k=\l$ corresponds to
the peak in $\langle c_{k}^{*} \rangle$ [Eq.~(\ref{eq:ckcl}] is
dominated by the $H P(n_{1}=k; t)$ term).  White regions in the
off-diagonal areas imply negative correlation between species counts
of large neighboring sizes.  In other words, whenever we observe a
species with 800 individuals in a simulation at any fixed time $t$ (at
QSS), we will probably not observe another species with 801 cells at
the same time. Grey areas that are farther away (such as $k=\l=600$)
represent transient states of the system and have near-zero
covariances.
%

\subsection{Other forms of global interactions}

Since global interactions across all species mediate the breakdown of
the mean-field approximation, we now investigate different forms of
regulation imposed through the functions $r(N)$ and $\mu(N)$.  To
explore how the ``stiffness'' of different total population
constraints affects the expected QSS species-count vector $\langle
c_{k}^{*}\rangle$, we consider a simple Hill-type birth function with
Hill coefficient 1:
\begin{eqnarray}
r(N) = \frac{p_2 K_2}{K_2+N}, \qquad \mu(N) = \mu_2.
\end{eqnarray}
This form imposes a ``softer" constraint on the total population $N$
than the logistic birth function.  In order to compare the results
with those of the logistic model in Subsection \ref{sec:res1}, we use
the same values of $\alpha$ and $H$ and use $\mu_2=\mu,~p_2=p$ and
$K_2 = K-N^*$ where $N^*$ is the QSS population size obtained from the
logistic model.

Another way to implement regulation is by keeping the birth
rate constant but allowing the death rate to be population-dependent:
\cite{parsons2008absorption}
\begin{equation}
r(N) = r_3, 
\qquad
\mu(N) = \mu_3\left(1+\frac{N}{K_3}\right).
\label{MUN}
\end{equation}
Again, we are interested in expected species counts near the same
$N^*$ as in Subsection \ref{sec:res1}, we set $K_3 = K,~ r_3 = r^*,~
\mu_3 = \mu^{*}/(1+\frac{N^*}{K_3})$, where $r^{*}$ and $\mu^{*}$ are
the QSS values of the birth and death rates used in the logistic
model.

Note that both the alternative regulation models, the Hill-type model
and the population-dependent death model, generate the same
steady-state rates $r^{*}$ and $\mu^{*}$ at the same QSS total
population size $N^*$ as in the logistic model.  Thus, we can compare
the expected species-counts from all three
models on the same footing.  Since the mean-field solution given in
Eq.~(\ref{Eq:mf}) depends only on $r^{*}$ and $\mu^{*}$, all three
models yield identical mean-field solutions $\langle
c_{k}^{*}\rangle$. Therefore, for not-too-small values of $\alpha$,
for which mean-field solutions are accurate, all three models yield
the same $\langle c_{k}^{*}\rangle$.

\begin{figure}[h!]
\centering
\includegraphics[width=3in]{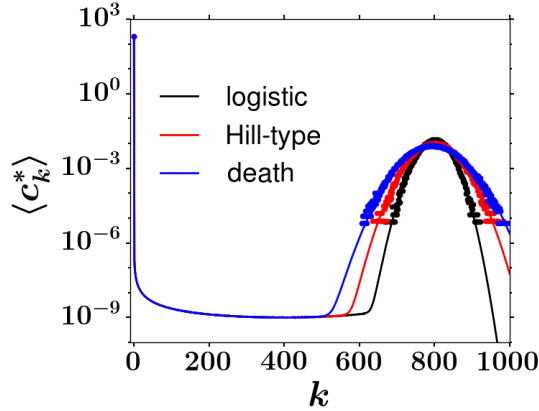}
\caption{Comparison of $\langle c_k^* \rangle$ across the three
  regulation models, logistic (narrowest, black), Hill-type
  (intermediate, red), and population-dependent death (widest,
  blue). Simulations and results from the weighted Moran model
  approximations are shown. Here, the immigration rate is $\alpha =
  10^{-8}$ where the mean-field approximation is invalid. The other
  parameters are $\mu_2=10,~p_2=20,~H=200, K=1600, r_{3}=r^{*}$, and
  $\mu_{3} = \mu^{*}$.}
%
\label{WIDTHS}
\end{figure}

However, for small $\alpha$, where the mean-field approximation breaks
down, we expect that the peak in $\langle c_{k}^{*}\rangle$ near $k =
N^*$ will be quantitatively different among the three models.  In
Fig.~\ref{WIDTHS}, we set $\alpha = 10^{-8}$ and plot the expected
species count (from simulations and our
weighted Moran model approximation) associated with each of the three
models.
%
%
%
Note that the peaks in $\langle c_{k}^{*}\rangle$ differ in
their widths.  In all examples, the underlying Moran models are
identical and the differences originate in the different
total-population distributions $P^*(N)$ across the three regulation
models, as illustrated by the different ``widths" of the peak near
$N^*$.  According to simulations and numerical solutions of our
weighted Moran model, the peak widths corresponding to each regulatory
model are ranked according to population-dependent death $>$ Hill-type $>$
logistic growth.

A wider peak in $\langle c_{k}^{*}\rangle$ can be associated with a
``softer" total population constraint. As long as $f(N)$ on the
right-hand side of Eq.~(\ref{DNDT}) is a differentiable near $N^*$, we
can define the regulatory ``stiffness'' by

\begin{equation}
|f_*'| = -f_*' \equiv \frac{\dd f(N)}{\dd N} \bigg|_{N^*} \in [0, +\infty).
\end{equation}
Note that $f_*' < 0$ as long as $N^*$ is a locally stable point.

The larger $|f_*'|$ is, the more likely the next event will be
``compensatory" (e.g. a new birth increases the chance for the next
event to be death).  This stiffness can be also thought of as the
curvature of a quadratic energy profile centered about $N^{*}$.  The
stiffnesses of our three examples (using $\alpha = 10^{-8}$) are $\vert
f_*'\vert = \vert p-\mu - \frac{2p}{K} N^*\vert = 10$ for logistic
birth, $\vert f_*'\vert = \vert p_2\frac{(N^*)^2+2K_2
  N^*}{(N^*+K_2)^2} - \mu_2\vert = 5$ for Hill-type regulation, and
$\vert f_*'\vert = \vert r_3 - \mu_3 - \frac{2\mu_3}{K_3}N^*\vert =
3.3$ for population-dependent death. These stiffness values are
consistent with the progression of peak widths shown in
Fig.~\ref{WIDTHS}.

We may extend our definition of the stiffness to cases where $f(N)$ is
not differentiable.  For example, the Moran model has an infinitely
``stiff" constraint ($f'_* = -\infty$) which ``forces" an immediately
death after a new birth. Nevertheless, as we have shown, even though a
regulated BDI model may have much less sensitivity than the Moran
model, the latter still provides insights on how such regulatory
effects can induce an expected species count that exhibits a
  peak at $N^*$.
%

\subsection{Energy landscape and phase transition in the species-count distribution}
\label{sec:landscape}
In this section, we provide an interpretation of the failure of the
mean-field equation [Eq.~(\ref{Eq:mf})] as a ``phase-transition" in
the statistics of populations.  The full high-dimensional BDI model in
QSS can be described by two processes: (1) evolution of individual
species fractions via a Moran model that is equivalent across
different regulation models, and (2) fluctuations of the total
population size according to a QSS distribution that depends on the
regulation model.  Since the failure of the mean-field approach arises
essentially from the emergence of a species that represents a large
fraction of the whole population, we focus on the contribution of the
Moran process.

A phase-transition can be conveniently visualized using a potential
energy landscape $\phi$ as is widely used in population genetics and
developmental biology \cite{wright1932roles,Waddington1957,
  Sherrington1997,Arnold2001,Ao2009,Orr2009,xu2014two}.  Its recent
development in the physics community has extended its application to
quantitative and systems biology \cite{Ao2004,Qian2006,Wang2008}.
Defined as a measure of ``generalized energy", its gradient indicates
the direction of evolution of the system and its minima (potential
wells) denote local stable states.


To simplify the math, we consider the $N \gg 1$ limit and use the
continuum limit of the Moran model to find a continuum energy
landscape $\phi(\{x_1,x_2,...\})$ such that 
$P^{*}_{\rm M}(\x) \propto e^{-\phi}$ satisfies Eq.~(\ref{eq:diffu})
in steady-state. The shape of $\phi$ across $\{x_1,x_2,...\}$ characterizes
the global stability of the model. Starting from the 1D version of
Eq.~(\ref{eq:diffu}) we have
\begin{equation}
A(x) = m^* \left( \frac{1}{H}-x \right), \quad
B(x) = x(1-x), 
\label{eq:diffu1D}
\end{equation}
which allow us to define the 1D energy function \cite{xu2014two}
\begin{align}
\phi(x) & \equiv -N^{*}\int^x \frac{A(y)}{B(y)} + \ln B(x) \nonumber \\
\: & = \left( 1-\frac{\alpha}{\mu^*} \right) \ln ( x ) + 
\left( 1-\frac{\alpha(H-1)}{\mu^*} \right) \ln(1-x) \nonumber \\
\: & \equiv \frac{1}{H-1} \left[ (H-1) - \frac{\alpha}{\alpha_{\rm c}} \right] \ln x+ 
\left( 1 - \frac{\alpha}{\alpha_{\rm c}} \right) \ln (1 - x).
\label{eq:phi}
\end{align}
Here, the parameter
\begin{equation}
\alpha_{\rm c} \equiv \frac{\mu^*}{H-1}, 
\label{eq:alpc}
\end{equation}
is a critical immigration rate that controls a ``phase transition."
Eq.~(\ref{eq:alpc}) is unambiguous when $\mu$ is constant. If the
regulation arises from a population-dependent rate $\mu(N)$ as in
Eq.~(\ref{MUN}), the critical immigration rate $\alpha_{\rm c}$ can be
approximated by self-consistently solving $\alpha_{\rm c} =
\mu(N^{*}(\alpha_{\rm c}))/(H-1)$.

When $P_{\rm M}^*(x)$ is normalisable, the energy function satisfies
$\phi(x) \propto -\ln P_{\rm M}^*$.
Since $\ln(0^{+}) \to -\infty$ and $\ln(1)=0$, the shape of $\phi(x)$
is determined by the signs of the coefficients $H-1 -
\frac{\alpha}{\alpha_{\rm c}}$ and $1 - \frac{\alpha}{\alpha_{\rm
    c}}$.  Assuming $\alpha_{\rm c} > 0$ (see Subsection
\ref{sec:resolvealphaH} for the special case $\alpha_{\rm c} = 0$),
different regimes of the model can be delineated

\begin{figure}[h!]
\centering
\:\hspace{-1mm}\includegraphics[width=6.4in]{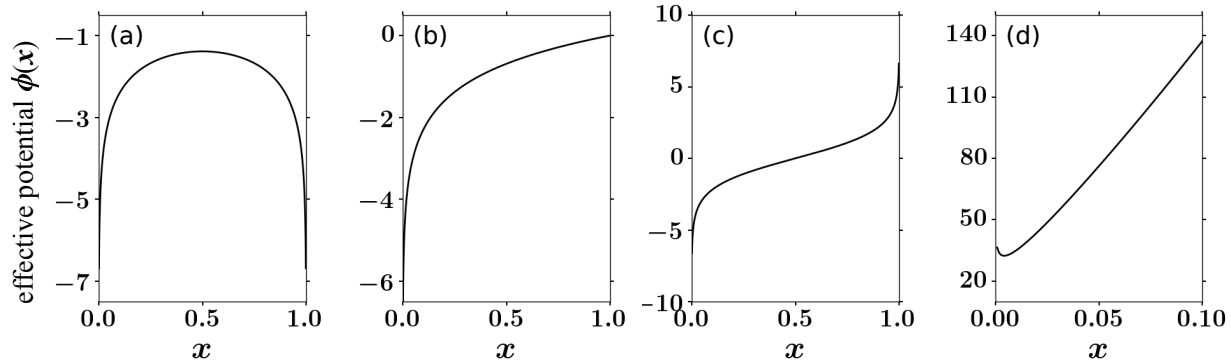}
\caption{Energy landscapes $\phi(x)$ as a function of $x = n/N^*$ for
  $\mu = 10,~ p = 20,~ K = 1600,~ H = 200$ and different values of
  $\alpha$.  (a) $\alpha = 10^{-8}$ corresponding to
  Fig.~\ref{fig:moran1d}(a), (b) $\alpha = \alpha_{\rm c}=10/199$, (c)
  $\alpha = 0.1$, and (d) $\alpha=60$. 
  The minimum at $x>0$
  corresponds to the peak in $\langle c_{k}^{*}\rangle$ arising at
  $k>1$.}
\label{Fig:phis}
\end{figure}

\begin{itemize}
\item
When $\alpha < \alpha_{\rm c}$, we have $\alpha < (H-1)\alpha_{\rm c}$
for $H\geq 2$.  Two infinite minima in $\phi(x)$ emerge; one at $x=0$
and one at $x=1$.  Associated with each minima is a basin of
attraction as shown in Fig.~\ref{Fig:phis}(a).  Even if all
species start with a small fraction $x_i \ll
1$, one of them can eventually come to dominate by crossing to the
attractive peak at $x=1$ causing a failure of the mean-field
description.  However, this transition is different from the usual
stochastically-driven ``escape" in statistical physics (see
Discussion).  When $\alpha$ is extremely small, the ``extinction"
state $x=0$ is approximately absorbing for each
species and the mean-field approximation fails
severely.

\item
When $\alpha = \alpha_{\rm c}$, $\alpha < (H-1)\alpha_{\rm c}$ for $H>
2$.  The potential $\phi(x) = {H-2\over H-1}\ln x$ is monotonic
and exhibits a global diverging minimum at $x=0$ and a global maximum
at 1.  The whole interval $[0,1]$ is a basin of attraction for $x=0$
as shown by Fig.~\ref{Fig:phis}(b).  The energy away from $x=0$ is
very flat and the severity of the failure of the mean-field approach
is sensitive to $\alpha$ when it is near $\alpha_{\rm c}$.

\item
When $(H-1)\alpha_{\rm c} \geq \alpha > \alpha_{\rm c}$, the potential
$\phi(x)$ has a diverging minimum at $x=0$ and a diverging maximum at
$x=1$ as shown in Fig.~\ref{Fig:phis}(c).  The mean-field approach is
accurate in this regime.

\item 
When $\alpha > (H-1)\alpha_{\rm c}$, there is a single finite minimum in
$\phi(x)$ appearing at $x_{\rm min} = \frac{\alpha - \alpha_{\rm
    c}(H-1)} {\alpha H - 2 \alpha_{\rm c} (H-1)}$, which is close to
$x=0$ when $H \gg 1$.  The potential has diverging maxima at both
$x=0$ and $x=1$ so the basin of attraction for $x_{\rm min}$ is the
whole $[0,1]$ interval as shown by Fig.~\ref{Fig:phis}(d).  The
mean-field approach is accurate in this case.
\end{itemize}

Physically, a small immigration rate $\alpha < \alpha_{\rm c}$ does
not allow a dominant species to be replaced by new ones. When $\alpha
> \alpha_{\rm c}$, the immigration frequency $\alpha H$ is larger that
the rate of coarsening of the species counts thereby filling the
system with new species and preventing any one species to
dominate. Here, $\langle c_{k}^{*}\rangle$ is monotonically
decreasing.  At even larger $\alpha > (H-1)\alpha_{\rm c}$,
immigration of each species is frequent enough that $\langle
c_{k}^{*}\rangle$ becomes very broad and again develops an interior
peak at $k \approx x_{\rm min}N^{*}$. The collapsing of $\langle
c_{k}^{*}\rangle$ into a single species is reminiscent of the collapse
of cluster size distributions in self-assembly under finite resources
\cite{NUCLEATION1}.

\subsection{Resolving the effects of $\alpha$ and $H$}
\label{sec:resolvealphaH}

The energy landscape formulation provides a general way to visualize
whether there is phase transition in the dynamics of an individual
species.  which can be applied to study how various parameter affect
the model and generalized to various models.  Equipped with the energy
landscape, we can now examine the species populations as the intrinsic
immigration rate $\alpha$ and the total number of species $H$ that can
immigrate varied, keeping the total immigration rate, $\alpha H$,
fixed.
Varying $\alpha$ and $H$ in this way will not change the dynamics of
the total population $N$ but will influence the dynamics of
populations of individual species.
%
%
This is readily shown by the different shapes of $\phi(x)$ as $\alpha$
and $H$ change.  For example, let $\mu^*=0.33$.  With $H=2,~\alpha=1$,
the landscape exhibits a ``most probable" species size maintained by
high per-species immigration rate.  However, if $H=200$ and
$\alpha=0.01$, each species has a low immigration rate.  The
associated landscape $\phi(x) = 0.97\ln x - 5\ln(1-x)$ exhibits a
unique potential well at $x=0$ as all species are driven small.
We can also consider the limit $H \rightarrow \infty$ while keeping
$\alpha H$ fixed.  This limit approximates naive T cell generation by
the thymus.  While total thymic output $\alpha H$ is finite, there are
theoretically $H > 10^{15}$ different species (T-cell receptor
sequences) that can be generated although only about $10^{6}-10^{8}$
different species survive \cite{lythe2016many}.
%
%
In any case, this large value of $H$ means that nearly every
immigration is from a new, unrepresented species and $k=0$ is an
absorbing boundary for all existing species.  Species labels keep
changing, but the distribution of $\langle c_k\rangle$ reaches a QSS.
The energy landscape becomes (taking $\alpha \rightarrow 0$ in
Eq.~(\ref{eq:phi})) $\phi(x) \to \ln x + \left( 1 - \frac{\alpha
  H}{\mu^*} \right) \ln (1 - x )$.  There is always an potential well
at $x=0$ while the dynamics near 1 depend on the sign of $\frac{\alpha
  H}{\mu^*}-1$.

Recall that if $\alpha < \alpha_{\rm c}$ the mean-field approximation
to $\langle c_{k}\rangle$ fails. From Eq.~(\ref{eq:alpc}), the
critical value $\alpha_{\rm c}$ increases as $H$ decreases rendering
the mean-field approximation invalid for a larger range of immigration
rates. When $\alpha_{\rm c}\to 0$ (e.g. realized when $\mu^* \to 0$),
Eq.~(\ref{eq:phi}) no longer has a valid form because birth would need
to be negative ($N^* > K$) in order to balance immigration.
Nonetheless, we can multiply the landscape function by a constant
$\mu^{*}$ without affecting its ability to qualitatively characterize and
classify the dynamics of the system.  We then take the limit
$\mu^*\rightarrow 0$ and get $\phi \propto - \alpha \ln ( x ) -
\alpha(H-1) \ln (1 - x )$, which always has a unique minimum between
$(0,1)$, corresponding to Fig.~\ref{Fig:phis}(d).

\section{Discussion and Summary}

In our analysis, the non-mean-field behavior of the expected clone
abundances are mediated by global regulation mechanisms that act
uniformly across all clones.  Such population-dependent interactions
break independence between clones, are difficult to model, and
consequently have been rarely discussed in the context of species
diversity \cite{volkov2005density}. Empirical studies have focused on
the small-to-intermediate range of $k$, where the distribution
$\langle c_{k}^{*}\rangle$ is well approximated by the mean-field
model (Eq.~(\ref{Eq:mf})) as seen in Figs.~\ref{fig:goyal} and
\ref{fig:moran1d}. In another study, Parsons {\it et al.}
\cite{parsons2008absorption} considered a neutral and quasi-neutral
birth-death model with carrying capacity but focussed on the mean
fixation time of any species rather than species counts. However, they
find that fluctuations in the total population do not affect fixation
times of neutral species which is consistent with our finding that a
fixed-population Moran process can be used to accurately construct the
fractions $\frac{n_1}{N}$ of any species in our BDI model. To our
knowledge, the failure of predicting a large-size clone by
Eq.~(\ref{Eq:mf}) has not been explicitly discussed in detail.

In many contexts such as stem or progenitor cells in a bone marrow
niche, or multiple species competing for common resources, the
observation of one or a few large clones or high population species is
often naturally attributed to selection (differences in growth or
death rates). This largest ``outlier'' clone can contain most of the
population and be biologically more important than all other smaller
clones/species in the organism/community.  Our results show that a
simpler mechanism may arise from slow immigration into a neutral
birth-death process with regulation, providing an initial ``null
hypothesis" for selection.  Otherwise, one may incorrectly argue that
the existence of such a singular outlier clone suggests a species
selection effect.

To quantitatively understand the regulated multispecies BDI process,
we showed that an often-used mean-field approximation captures the
expected steady-state species counts at low populations, but
completely misses a possible peak in the species abundance near the
population supported by a general regulated birth-death-immigration
process.  This peak arises only when the immigration rate decreases
below a threshold value.

To develop a theory that approximates the clone abundance distribution
accurately in all parameter regimes, we then mapped the $q^{\rm th}$
moment of the species abundance distribution
$c_k$ to a $(q+1)$-dimensional cell-count BDI model, which was then
approximated by weighting over $q$-dimensional Moran models of
different system size.  The expected distribution and covariances of
species counts were accurately calculated in
parameter regimes in which the mean-field approximations break down.
By exploiting the concept of energy landscapes, we analytically
describe a phase transition in the dynamics which explains the
failure of the mean-field approach in the original model.  Our
analysis shows that global (inter-species) carrying capacity,
when combined with a random sampling mechanism, generates a
genetic-drift-like effect \cite{ewens2012mathematical} in a Moran
model that ultimately destroys the universal power-law distribution of
$c_k$.

In Eq.~(\ref{Eq:truemean}), dynamics of any $\langle c_k^{*} \rangle$
are controlled by $r(N)$, where $N \equiv \sum_\l \l c_\l$.  It
involves contributions from all clone populations $\l=0,1,2,...,k,...$
Recall that in many classical scenarios the relative strengths of
these effects on the $k^{\rm th}$ component decay with distance $|
k-\l |$. For example, in the constant-rate BDI model, only $c_{k\pm
  1}$ and $c_k$ affect the dynamics of $c_k$.
%
%
Here, however, the contribution from $c_\l$ is proportional to the
index $\l$ itself instead of on $| k-\l |$.  This is a type of
``long-range" interaction or long-distance coupling arise in
theoretically challenging contexts in different areas
\cite{morchio1987mathematical,takayama2012cooperative,sanyal2012long,
  chen2016long}.  Thus, the structure of $\langle c_k^{*} \rangle$
according to Eq.~(\ref{Eq:truemean}) can no longer be approximated by
a simple monotonic form as is shown in Fig.~(\ref{fig:moran2d}) where
correlations between large $k$ and its neighboring states $k\pm1$ are
negative.


To effectively find higher moments of species counts in QSS,
higher-dimensional Moran models can be used to construct the related
QSS cell-count distributions.  Diffusion approximations to
high-dimensional Moran models provide convenient analytic-form
steady-state distributions
\cite{ewens1963numerical,aalto1989moran,baxter2007exact}.  However,
the boundary values of $P^*_{\rm M}(\x)$ in the diffusion
approximation may not accurately approximate those from the discrete
Moran model, especially in higher dimensions. For example, when $N^{*}
m^{*} \ll 1$ in Eq.~(\ref{Eq:P-moran-conti-solu}), $P_{\rm M}(x_i=0) =
+\infty$ but $P_{\rm M}(x_i=\frac{1}{N}) \approx 0$.  Near the
boundary, $P_{\rm M}(0<x<\frac{1}{N})$ generally changes in a highly
non-linear fashion.  Only with extremely large $N$ do the probability
distributions of the discrete and continuous Moran models match well
\cite{doering2005extinction,kessler2007extinction}.  Nevertheless, our
result in Fig.~\ref{fig:moran2d} is accurate because the region of
interest is far away from the boundaries. Moreover, the second moment
$\langle (c_{k}^{*})^{2} \rangle$ in Eq.~(\ref{eq:ckcl}) turns out to
be dominated by the first-moment term $H\langle c_k^{*} \rangle$ which
was calculated based on the exact discrete solution in
Eq.~(\ref{eq:moran1d}).

%
%

%
%

%
It is worth noting that the \textit{initial} establishment of the
large clone $i$, denoted by the transition $x_i\approx 0 \rightarrow
x_i\approx 1$, is different from traditional scenarios where clone $i$
randomly crosses the energy barrier near $x=0.5$ and ``escapes" to the
other attractive basin.  Here, the potential energy profile
corresponds to QSS in which there are many different clones $j$
starting with small fractions $x_j\approx 0$. One of these small
clones eventually replaces the dominating clone $i$ ($x_j\approx 1$).  The waiting
time for such replacement event was obtained by \cite{xu2014two} as
$T_{\rm r} \sim \mathcal{O}(\frac{N^*}{\alpha(H-1)})$, a much longer
time than the waiting time $T_2 \sim \mathcal{O}(\frac{N^*}{\mu^*})$
(see Appendix \ref{Ap:multi}) for the establishment of the first
dominant clone in our BDI regulated model under $\alpha(H-1)\ll \mu^*$.

Future improvements to our analysis include more accurately determining
steady-state solutions of the higher dimensional Moran models,
especially near the boundaries and extending our approaches to
time-dependent approximations. To better distinguish our neutral
mechanism from true selection, a careful analysis of heterogeneous
populations should be explored to determine how random dominance from
neutral regulation might be balanced by selection in the form of
heterogeneous growth, death, and immigration parameters.

\section{Acknowledgments}
This work was supported in part by grants from the NSF (DMS-1516675
and DMS-1814364) and the Army Research Office (W911NF-18-1-0345).

\bibliographystyle{unsrt}
\bibliography{references_ck}

\newpage
\appendix
\section*{Mathematical Appendices}

\section{Cell-count Master equation for $P(n_1,...,n_i,...,n_{H}; t)$}
\label{appendix:pn}
The high-dimensional master equation obeyed by the full multispecies
distribution reads
\begin{eqnarray}
&&
{\partial  P(\n;t)  \over \partial t} =  
\alpha \sum_{i=1}^{H}   P(n_1,...,n_{i-1}, n_i - 1, n_{i+1},...,n_{H})  -  P(\n) ] 
\notag \\ && \qquad
+ \sum_{i=1}^{H} \left[r(N-1) (n_i - 1) P(n_1,..., n_i - 1,...,n_{H}) 
- r(N) n_{i} P(\n)\right] 
\notag \\ && \qquad
+ \sum_{i=1}^{H} \left[\mu(N+1) (n_{i}+1)  P(n_1,..., n_i + 1,...,n_{H}) 
- \mu(N) n_i P(\n)\right],
\label{mastereqn:pn} 
\end{eqnarray}
where $N\equiv \sum_{i=1}^{H} n_i$.

\section{Dynamical equations for $\langle c_{k}(t) \rangle$}
\label{ap:ckreal}

Define $P(\c; t)$ as the probability of observing the configuration
$\c = \{c_0, c_1, c_2, ... \}$ at a specific time $t$.  Under constant
immigration and population-regulated birth and death rates, the
evolution of the full probability distribution satisfies the master
equation
\begin{eqnarray}
&& \frac{\partial P(\c; t)}{\partial t} 
= - \sum_{k=0}^{\infty} [\alpha + (\mu(N)+r(N))k] c_k 
P(\c) 
\notag \\ && \qquad\qquad\qquad
+\sum_{k=0}^{\infty} (c_{k+1}+1) (k+1)\mu(N+1) P(\{...,c_{k}-1,c_{k+1}+1,...\}) 
\notag \\ && \qquad\qquad\qquad
+ \sum_{k=0}^{\infty} (c_{k}+1) (\alpha + k r(N-1)) P(\{...,c_{k}+1,c_{k+1}-1,...\}).
\label{eq:Pc}
\end{eqnarray}
Without loss of generality, 
let us assume constant $\mu,~\alpha$ but regulated $r=r(N)=r(\sum_{k=1}^\infty k c_k)$. 
%
%
The expected clone count is
\begin{eqnarray}
&&
\langle c_\l(t) \rangle = \sum_{c_\l =0}^{H} c_\l P(c_\l; t)  
= \sum^H_{c_0=0} \sum^H_{c_1=0}  ... 
\sum_{c_k=0}^H ... ~ c_\l P(c_0,c_1,...,c_{\l-1},c_\l,c_{\l+1},...; t). 
\label{Eq:cl-def}
\end{eqnarray}
%
%
%
Substituting Eq.~(\ref{eq:Pc}) into Eq.~(\ref{Eq:cl-def}), we obtain
\begin{eqnarray}
&&
\frac{\dd \langle c_\l(t) \rangle}{\dd t} = \sum^H_{c_0=0}
\sum^H_{c_1=0}  ... \sum_{c_k=0}^H ... ~ c_\l 
\frac{\partial P(c_0,c_1,...,c_{k-1},c_k,c_{k+1},...;t)}{\partial t}
\notag \\ &&
= 
\sum^H_{c_0=0}
\sum^H_{c_1=0}  ... \sum_{c_k=0}^H ... ~ c_\l 
\bigg\{ \sum_{k=0}^{\infty} -[\alpha + (\mu+r(N))k] c_k 
P(c_0,c_1,...,c_{k-1},c_k,c_{k+1},...)
\notag \\ && \qquad 
+\sum_{k=0}^{\infty} (c_{k+1}+1) [(k+1)\mu] P(c_0,c_1,...,c_{k-1},c_{k}-1,c_{k+1}+1,...) 
\notag \\ && \qquad
+ \sum_{k=0}^{\infty} (c_{k}+1) [\alpha + k r(N-1)] P(c_0,c_1,...,c_{k-1},c_{k}+1,c_{k+1}-1,...)
\bigg\}.
\label{tmp:ckall}
\end{eqnarray}
By collecting only terms in Eq.~(\ref{tmp:ckall}) that involve $r(N)$, we obtain two summations
\begin{eqnarray}
&&
S_1 + S_2 \equiv 
- \sum^H_{c_0=0}
\sum^H_{c_1=0}  ... \sum_{c_k=0}^H ... ~ c_\l
 \sum_{k=0}^{\infty} r(N) kc_k P(c_0,c_1,...,c_{k-1},c_k,c_{k+1},...)
 \notag \\ && \qquad 
 +
\sum^H_{c_0=0}
\sum^H_{c_1=0}  ... \sum_{c_k=0}^H ... ~ c_\l 
\sum_{k=0}^{\infty} r(N-1) k(c_{k}+1) P(c_0,c_1,...,c_{k-1},c_{k}+1,c_{k+1}-1,...).
\nonumber
\end{eqnarray}
Consider the contribution of the $k^{\rm th}$ terms in both
summations:

\begin{itemize}
\item When $k<\l-1$ or $k\geq \l+1$, the $k^{\rm th}$ term of $S_1$ becomes
\begin{eqnarray}
- \sum^H_{c_0=0}
\sum^H_{c_1=0}  ... \sum_{c_k=0}^H ... ~ c_2
r(N) (k-1) c_{k-1} P(c_0,c_1,c_2,...,c_k,...) 
 \end{eqnarray}
and $k^{\rm th}$ term of $S_2$ becomes
\begin{eqnarray}
&&
 \sum^H_{c_0=0} ... \sum_{c_{k-1}=0}^H \sum_{c_k=0}^H ... ~ c_\l 
r(N-1) (k-1) (c_{k-1}+1) P(c_0,c_1,...,c_{k-1}+1,c_k-1,...) 
\notag \\ &&
= \sum^H_{c_0=0} ... \sum_{c_{k-1}=0}^H \sum_{c_k=0}^{H-1} ... ~ c_\l 
r(N) (k-1) c_{k-1} P(c_0,c_1,c_2,...,c_{k-1},c_k,...) 
\notag \\ &&
= \sum^H_{c_0=0} ... \sum_{c_{k-1}=0}^H \sum_{c_k=0}^{H} ... ~ c_\l 
r(N) (k-1) c_{k-1} P(c_0,c_1,c_2,...,c_{k-1},c_k,...). 
\end{eqnarray}
The last equality holds since $P(c_k=H)=0$ due to the constraint 
that if $c_k=H$, then 
all other $c_{m\neq k}=0$. 

\item When $k=\l-1$, the $k^{\rm th}$ term of $S_1$ is
\begin{eqnarray}
- \sum^H_{c_0=0} ... \sum_{c_{\l-1}=0}^H \sum_{c_\l=0}^H ... ~ c_\l 
r(N) (\l-1) c_{\l-1} P(c_0,c_1,c_2,...,c_k,...)
 \end{eqnarray}
and the $k^{\rm th}$ term of $S_2$ is
\begin{eqnarray}
&&
\sum^H_{c_0=0} ... \sum_{c_{\l-1}=0}^H \sum_{c_\l=0}^H ... ~ c_\l 
r(N-1) (\l-1) (c_{\l-1}+1) P(c_0,c_{\l-1}+1,c_{\l}-1,c_3,...,c_k,...)
\notag \\ && \qquad 
= \sum^H_{c_0=0} ... \sum_{c_{\l-1}=0}^H \sum_{c_\l=0}^{H-1} ... ~ (c_\l+1) 
r(N) c_{\l-1} P(c_0,c_1,c_2,c_3,...,c_k,...)
\notag \\ && \qquad
= \sum^H_{c_0=0} ... \sum_{c_{\l-1}=0}^H \sum_{c_\l=0}^{H} ... ~ (c_\l+1) 
r(N) c_{\l-1} P(c_0,c_1,c_2,c_3,...,c_k,...).
\end{eqnarray}

The two terms sum to 
\begin{eqnarray}
\sum^H_{c_0=0}
\sum^{H}_{c_1=0}  \sum_{c_2=0}^H ... \sum_{c_k=0}^H ... ~ 
r(N) c_{\l-1} P(c_0,c_1,c_2,...,c_k,...) = \langle r(N) c_{\l-1} \rangle.
\end{eqnarray}

\item When $k=\l$, the $k^{\rm th}$ term of $S_1$ is
\begin{eqnarray}
- \sum^H_{c_0=0} ... \sum_{c_{\l-1}=0}^H \sum_{c_\l=0}^H ... ~ c_\l 
r(N) \l c_\l P(c_0,c_1,c_2,...,c_k,...)
 \end{eqnarray}
while the $k^{\rm th}$ term of $S_2$ is
\begin{eqnarray}
&&
\sum^H_{c_0=0} ... \sum_{c_{\l}=0}^H \sum_{c_{\l+1}=0}^H ... ~ c_\l 
r(N-1) \l (c_{\l}+1) P(c_0,c_1,c_\l+1,c_{\l+1}-1,...,c_k,...)
\notag \\ && \qquad 
= \sum^H_{c_0=0} ... \sum_{c_{\l}=0}^H \sum_{c_{\l+1}=0}^{H-1} ... ~ (c_\l-1) 
r(N) \l c_{\l} P(c_0,c_1,c_2,...,c_k,...)
\notag \\ && \qquad
= \sum^H_{c_0=0} ... \sum_{c_{\l}=0}^H \sum_{c_{\l+1}=0}^H ... ~ (c_\l-1) 
r(N) \l c_{\l} P(c_0,c_1,c_2,...,c_k,...).
\end{eqnarray}
These two terms sum to 
\begin{eqnarray}
\sum^H_{c_0=0} ... \sum_{c_{\l}=0}^H \sum_{c_{\l+1}=0}^H ...  
r(N) \l (-c_{\l}) P(c_0,c_1,c_2,...,c_k,...) = -\l \langle r(N) c_\l \rangle.
\end{eqnarray}
\end{itemize}
Summarizing, terms that involve $r(N)$ in Eq.~(\ref{tmp:ckall}) are simplified as 
$(\l-1)\langle r(N) c_{\l-1} \rangle - \l \langle r(N) c_\l \rangle$. 
Terms involving $\alpha$ and $\mu$ can be similarly obtained if they are regulated by $N$.
Together, Eq.~(\ref{tmp:ckall}) becomes
\begin{equation}
\displaystyle {\dd \langle c_\l \rangle \over \dd t} = 
\alpha (\langle c_{\l-1} \rangle - \langle c_\l \rangle) +  
\langle r(N) \left[(\l-1) c_{\l-1} - \l  c_\l \right]\rangle
+ \langle  \mu(N) \left[(\l+1) c_{\l+1} - \l c_\l \right] \rangle.
\end{equation}
%

\section{Multi-timescale dynamics of $N(t)$ and $c_k(t)$}
\label{Ap:multi}
For simplicity, we first discuss the model with no immigration ($\alpha = 0$) and 
a large carrying capacity $K$. 
In this limit, $N^* = (1-\frac{\mu}{p})K \sim \mathcal{O}(K^{-1})$. 
The deterministic Eq.~(\ref{DNDT}) gives quite a good approximation
for the typical dynamics for $N$ in its first phase of evolution as
$N(t)$ quickly approaches its QSS value $\langle N^*\rangle$.  To
estimate this timescale, one can integrate $\frac{\dd N}{\dd t}$ in
Eq.~(\ref{eq:logi-setup}) under $\alpha=0$ to find $\langle
N(t)\rangle = \frac{\langle N^*\rangle N_0}{N_0 + e^{-(p-\mu)t}
  (\langle N^*\rangle - N_0)}$.  Thus $N$ approaches $\langle
N^*\rangle$ in a characteristic timescale
$\mathcal{O}(\frac{1}{p-\mu})$.

As $N$ approaches $\langle N^{*}\rangle$, $r(N)$ also approaches
$r(\langle N^{*}\rangle) \equiv r^*$, allowing $\langle c_k \rangle$
to approach its QSS value $\langle c_{k}^{*}\rangle$ which has a
high peak at $k=0$ and a small peak at $k \approx \langle N^*\rangle$.
Although this peak is small, the number of individuals in this 
clone, $k c_{k\approx \langle N^{*}\rangle}$ can comprise nearly the entire population.
This configuration is associated with a single
large-size clone that persists after the disappearance of all
other $H-1$ clones. 
If we define the number of living clones (or ``species richness")
\begin{equation}
R \equiv \sum_{k=1}^{\infty} c_k,
\end{equation} 
this ``coarsening'' or ``fixation'' process \cite{ewens2012mathematical} decreases $R$
from its initial value $H$ to 1 in finite time.
We define the waiting time for such a fixation to take place as $T_1$.
Since fixation is most relevant to changes of fractions of clones, we
study the problem in a Moran model.
In a standard textbook such as \cite{ewens2012mathematical}, 
the mean time for the $i^{\rm th}$ clone to fix (conditioned on 
its fixation) is
$T_{\rm fix}(i) \approx - \frac{N^*}{\mu}
\frac{N(0)-n_{i}(0)}{n_{i}(0)} \ln \left[ 1 - \frac{n_i(0)}{N(0)}
  \right]$.  The expected time until any arbitrary clone's fixation is
then calculated by averaging each clone's fixation times over its 
its probability of fixation ($P_{\rm fix}(i) = x_{i}(0)$) as $T_{\rm c}
= \sum_i T_{\rm fix}(i) P_{\rm fix}(i) \approx \frac{N^*}{\mu}$.

The last clone, which is just the total population, is stabilized to
$N^*$ by the regulatory effect of $f(N)$.
%
It fluctuates around $N^*$ for an exponentially long time. 
The variance (``width") of such fluctuation near $N^*$ can be 
calculated by invoking the full stochastic model  
Eq.~(\ref{eq:PN}) which leads to the 
solution $P^*(N)$ in Eq.~(\ref{Eq:PNsolu}). 
The fact that $P^*(0)\neq 0$ (although it is typically exponentially
small) allows for a finite probability that $N$ may incur a large
deviation to the absorbing boundary $N=0$, resulting in extinction of
the total population \cite{kessler2007extinction}.
%
%
The expected time to extinction of the total population is
$T_{\rm ext} = \sum_{m=1}^n a_m$ where $a_m = \frac{1}{\mu m} +
\sum_{j=1}^{\infty} \frac{1}{\mu (m+j)} \prod_{i=1}^{j}
\frac{r_{m+i-1}}{\mu}$ \cite{doering2005extinction}. 
The asymptotic approximation $T_{\rm ext} \sim \mathcal{O}(e^{\langle
  N^{*}\rangle})$ indicate a very long timescale for extinction, well
after QSS limit of $\langle c_{k}^{*}\rangle$ is approached.


\begin{figure}[h!]
\centering
\includegraphics[scale=0.66]{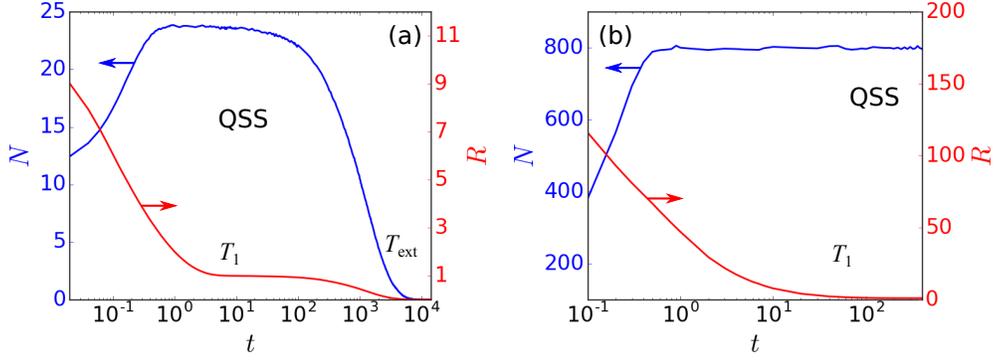}
\caption{Simulations of the multi-timescale dynamics of a small (a)
  and a large (b) system.  Common parameters are $\mu=10,~p=20$.
  Different parameters are $K=50,~H=11,~\alpha=0$ for (a) and
  $K=1600,~H=200,~\alpha=10^{-8}$ for (b).}
\label{fig:multitime}
\end{figure}

In Fig.~\ref{fig:multitime}, we plot simulations of the dynamics of
both $N$ and $R$ under two different sets of parameters. Values of
$\alpha$ are set to be extremely small or 0.
Fig.~\ref{fig:multitime}(a) shows that $N$ reaches $N^*\approx 25$
within 1 unit of time and remains stable over approximately $10^2$
before extinction. For $\alpha = 0$, we can identify a QSS within the
time period $T_1 < t < T_{\rm ext}$.  Fig.~\ref{fig:multitime}(b)
incorporates a small immigration $\alpha = 10^{-8}$ so that $N=0$ is
technically no longer an absorbing boundary.  Nonetheless, for
extremely small $\alpha H T_{\rm ext} \ll 1$, the dynamics are similar
to the $\alpha = 0$ case (Fig.~\ref{Fig:phis}(a)) since the
inter-immigration times $1/(\alpha H)$ are longer than the
extinction time of the whole population.

%

%
%
%

\section{Moments} 
\label{Ap:Moments}

The first moment of $\c$ is readily obtained by invoking its 
definition in Eq.~(\ref{CKDEF}) as 
\begin{eqnarray}
&&
  \langle c_k \rangle = \sum_{\n} [\mathds{1}(n_1,k) + \mathds{1}(n_2,k) + ...
    + \mathds{1}(n_H,k)] P(\n)
= H \sum_{\n} \mathds{1}(n_1,k) P(\n)
= H P(k)
\notag
\end{eqnarray}

The second moment, when $k \neq \l$, is obtained as
\begin{eqnarray}
&&
\langle c_k c_l \rangle = \sum_{\n} [\mathds{1}(n_1,k) + ... + \mathds{1}(n_H,k)]
[\mathds{1}(n_1,\l)+ ... + \mathds{1}(n_H,\l)] P(\n)
\notag \\ && \qquad
= \sum_{\n} \sum_{i=1}^H \mathds{1}(n_i, k) [\mathds{1}(n_1,\l) + ... + \mathds{1}(n_H,\l)] P(\n)
\notag \\ && \qquad
= H \sum_{\n} \sum_{j\neq 1} \mathds{1}(n_1, k) \mathds{1}(n_j,\l) P(\n)
= H (H-1) \sum_{\n} \mathds{1}(n_1, k) \mathds{1}(n_2,\l) P(\n)
\notag \\ && \qquad
= H (H-1) P(k, \l).
\notag
\end{eqnarray}
When $k = \l$, we have 
\begin{eqnarray}
&&
\langle c_k c_k \rangle = \sum_{\n} [\mathds{1}(n_1,k) + ... + \mathds{1}(n_H,k)]
[\mathds{1}(n_1,k)+ ... + \mathds{1}(n_H,k)] P(\n)
\notag \\ && \qquad
= H (H-1) P(k, k) + H\sum_{\n} \mathds{1}(n_1,k) \mathds{1}(n_1,k) P(\n)
\notag \\ && \qquad
= H (H-1) P(k, k) + H P(k). 
\notag
\end{eqnarray}

The third moment, when $k \neq \l \neq m$, is obtained as
\begin{eqnarray}
&&
\langle c_k c_\l c_m \rangle = \sum_{\n} [\mathds{1}(n_1,k) + ... + \mathds{1}(n_H,k)]
[\mathds{1}(n_1,\l) + ... + \mathds{1}(n_H,\l)] [\mathds{1}(n_1,m) + ... + \mathds{1}(n_H,m)] P(\n)
\notag \\ && \qquad
= \sum_{\n} \sum_{i=1}^H \mathds{1}(n_i, k)
[\mathds{1}(n_1,\l) + ... + \mathds{1}(n_H,\l)]
[\mathds{1}(n_1,m) + ... + \mathds{1}(n_H,m)] P(\n)
\notag \\ && \qquad
= H \sum_{\n}\mathds{1}(n_1, k) \sum_{i \neq 1} \mathds{1}(n_i,\l)
\sum_{j\neq 1,i} \mathds{1}(n_j,m) P(\n)
\notag \\ && \qquad
= H (H-1) (H-2) P(k, \l, m).
\notag
\end{eqnarray}
When $k = \l \neq m$, we have
\begin{eqnarray}
&&
\langle c_k^2 c_m \rangle = H (H-1) (H-2) P(k,k,m)
+ \sum_{\n} [\mathds{1}(n_1,k) + ... + \mathds{1}(n_H,k)] [\mathds{1}(n_1,m) + ...
  +\mathds{1}(n_H,m)]
\notag \\ && \qquad
= H (H-1) (H-2) P(k,k,m) + H(H-1) P(k,m).
\notag
\end{eqnarray}
And finally when $k = \l = m$, we obtain
\begin{eqnarray}
&&
\langle c_k^3 \rangle = 
H (H-1) (H-2) P(k,k,k) + H(H-1) P(k,k) + \sum_{\n} [\mathds{1}(n_1,k) + ...
  + \mathds{1}(n_H,k)]
\notag \\ && \qquad
= H (H-1) (H-2) P(k,k,k) + H(H-1) P(k,k) + H P(k).
\notag
\end{eqnarray}

\section{Diffusion approximation by the Taylor expansion}
\label{ap:diffusion}
For notational simplicity, we replace $x_1$ with continuous variables
$x$ and neglect the subscript ``M" in the Moran model probability
$P_{\rm M}$ in the rest of this subsection.  Letting $\ve =
\frac{1}{N^*} \rightarrow 0$ ($N^* \rightarrow \infty$) in
Eq.~(\ref{eq:Moran-master}), we expand the transition rates to 
second order in $\ve$:
\begin{eqnarray}
&&
\omega_{12} (x-\ve) P (x-\ve) \approx
(\omega_{12} P) -  \ve(\omega_{12} P)' + \frac{\ve^2}{2} (\omega_{12} P)'',
\\ &&
\omega_{21} (x+\ve) P (x+\ve) \approx
(\omega_{21} P) +  \ve(\omega_{21} P)' + \frac{\ve^2}{2} (\omega_{21} P)''.
\end{eqnarray}
Substituting them into Eq.~(\ref{eq:Moran-master}),
considering $\omega_{12}(x) = \alpha(1-x) + r^* N^* x(1-x),~
\omega_{21}(x) = \alpha(H-1)x + r^*N^* x(1-x)$ in Eq.~(\ref{eq:1d_fb}), 
and canceling out terms, we obtain (when $\alpha H \ll r^* N^*$)
\begin{eqnarray}
&&
{\rm RHS} \approx - \ve [(\omega_{12} - \omega_{21}) P]' 
+ \frac{\ve^2}{2} [(\omega_{12} + \omega_{21}) P]'' 
\notag \\ && \qquad
= - \frac{\alpha H}{N^*} \frac{\partial}{\partial x}\left( \frac{1}{H}-x \right)P + 
\frac{1}{2(N^*)^2} \frac{\partial^2}{\partial x^2} [\alpha(1-x)+\alpha(H-1)x+2r^*N^* x(1-x)]P
\notag \\ && \qquad \approx 
- \frac{\alpha H}{N^*} \frac{\partial}{\partial x}\left( \frac{1}{H}-x \right)P + 
\frac{r^* N^*}{(N^*)^2} \frac{\partial^2}{\partial x^2} x(1-x)P
\notag \\ && \qquad \approx 
\mu^* N^*
\left[ -\frac{1}{N^*} \frac{\partial}{\partial x} m^* \left( \frac{1}{H} - x \right)P_{\rm M}(x) + 
\frac{1}{(N^*)^2} \frac{\partial^2}{\partial x^2} x(1-x) P_{\rm M}(x) \right]
\end{eqnarray}
where $m^* = \frac{\alpha H}{\mu^* N^*}$ is the fraction of birth that comes from immigration.

For the 2D Moran model, we have 
\begin{eqnarray}
&&
\frac{\partial P(x_1,x_2)}{\partial t}
= (\omega_{21} P)(x_1+\ve, x_2-\ve) + (\omega_{31}) P(x_1+\ve, x_2) 
+ (\omega_{12} P)(x_1-\ve, x_2+\ve)
\notag \\ && \qquad
+ (\omega_{32} P)(x_1, x_2+\ve) + (\omega_{13} P)(x_1-\ve, x_2) + (\omega_{23} P)(x_1, x_2-\ve)
\notag \\ && \qquad
- [(\omega_{21}+\omega_{31}+\omega_{21}+\omega_{32}+\omega_{13}+\omega_{23}) P](x_1, x_2)
\label{eq:Moran-master-2D}
\end{eqnarray}
where
\begin{eqnarray}
&&
\omega_{21} = \alpha x_1 + r^* N^* x_2 x_1,\qquad \omega_{31} = \alpha (H-2) x_1 + r^* N^* x_3 x_1,  
\label{eq:omega1s}
\\ &&
\omega_{12} = \alpha x_2 + r^* N^* x_1 x_2,\qquad \omega_{32} = \alpha (H-2) x_2 + r^* N^* x_3 x_2,
\label{eq:omega2s}
\\ &&
\omega_{13} = \alpha x_3 + r^*  N^* x_1 x_3,\qquad \omega_{23} = \alpha x_2 + r^* N^* x_3 x_2.
\label{eq:omega3s}
\end{eqnarray}
Invoking the 2D Taylor expansion on Eq.~(\ref{eq:Moran-master-2D}), 
we obtain terms like
\begin{eqnarray} 
&&
(\omega_{21} P)(x_1+\ve, x_2-\ve) \approx 
(\omega_{21} P) + \ve \left[ \frac{\partial (\omega_{21} P)}{\partial x_1} 
- \frac{\partial (\omega_{21} P)}{\partial x_2} \right] 
\notag \\ && \qquad\qquad\qquad\qquad\qquad\qquad
+ \frac{\ve^2}{2} \left[ \frac{\partial^2 (\omega_{21} P)}{\partial x_1^2}
- 2\frac{\partial (\omega_{21} P)}{\partial x_1} \frac{\partial (\omega_{21} P)}{\partial x_2} 
+ \frac{\partial^2 (\omega_{21} P)}{\partial x_2^2} \right].  
\notag
\end{eqnarray} 
%
The right-hand side of Eq.~(\ref{eq:Moran-master-2D}) is thus approximated by
\begin{eqnarray}
&&
{\rm RHS} \approx 
\ve \left[ \frac{\partial (\omega_{21} P)}{\partial x_1} 
- \frac{\partial (\omega_{21} P)}{\partial x_2} \right] 
+ \frac{\ve^2}{2} \left[ \frac{\partial^2 (\omega_{21} P)}{\partial x_1^2}
- 2\frac{\partial^2 (\omega_{21} P)}{\partial x_1 \partial x_2} 
+ \frac{\partial^2 (\omega_{21} P)}{\partial x_2^2} \right]
\notag \\ && \qquad\qquad
+ \left[\ve\frac{\partial (\omega_{31} P)}{\partial x_1} 
+ \frac{\ve^2}{2} \frac{\partial^2 (\omega_{31} P)}{\partial x_1^2}
 \right]
+ \left[\ve\frac{\partial (\omega_{32} P)}{\partial x_2} 
+ \frac{\ve^2}{2} \frac{\partial^2 (\omega_{32} P)}{\partial x_2^2}
\right]
\notag \\ && \qquad\qquad
+ \ve\left[ -\frac{\partial (\omega_{12} P)}{\partial x_1} 
+ \frac{\partial (\omega_{12} P)}{\partial x_2} \right] 
+ \frac{\ve^2}{2} \left[ \frac{\partial^2 (\omega_{12} P)}{\partial x_1^2}
- 2\frac{\partial^2 (\omega_{12} P)}{\partial x_1 \partial x_2}
+ \frac{\partial^2 (\omega_{12} P)}{\partial x_2^2} \right]
\notag \\ && \qquad\qquad
+ \left[ -\ve\frac{\partial (\omega_{13} P)}{\partial x_1} 
+ \frac{\ve^2}{2} \frac{\partial^2 (\omega_{13} P)}{\partial x_1^2}
\right]
+ \left[ -\ve\frac{\partial (\omega_{23} P)}{\partial x_2} 
+ \frac{\ve^2}{2} \frac{\partial^2 (\omega_{23} P)}{\partial x_2^2}
\right]
\notag \\ && \qquad
= \ve\left[ \frac{\partial}{\partial x_1} (\omega_{21}+\omega_{31}-\omega_{12}-\omega_{13}) P 
+ \frac{\partial}{\partial x_2} (\omega_{12}+\omega_{32}-\omega_{21}-\omega_{23}) P \right]
\notag \\ && 
+ \frac{\ve^2}{2} \left[
\frac{\partial^2}{\partial x_1^2} (\omega_{21}+\omega_{31}+\omega_{12}+\omega_{13}) P
+ \frac{\partial^2}{\partial x_2^2} (\omega_{12}+\omega_{32}+\omega_{21}+\omega_{23}) P
- 2\frac{\partial^2}{\partial x_1 \partial x_2} (\omega_{12} + \omega_{21})P
\right]
\notag \\ && \qquad
= 
\mu^* N^* \left[ -\frac{1}{N^*} \sum_{i=1}^{2}  \frac{\partial A_i(\x) P(\x)}{\partial x_i}
+ \frac{1}{(N^*)^2} \sum_{i=1}^{2} \sum_{j=1}^{2} \frac{\partial^2 B_{ij}(\x)P(\x)}{\partial x_i \partial x_j}
\right]
\label{eq:diffu2Dderive}
\end{eqnarray}
where 
\begin{eqnarray}
&&
A_i(\x) = \sum_{j=1}^{2} m^* (Q_i - x_i), ~~  
B_{ii}(\x) = x_i (1-x_i), ~~ B_{ij}(\x)=-x_ix_j  ~ (i\neq j). 
\end{eqnarray}
The last step of Eq.~(\ref{eq:diffu2Dderive}) involves calculations based on
Eqs.~(\ref{eq:omega1s}-\ref{eq:omega3s}) and the assumption $m^*\ll 1$. 
For example,  
\begin{eqnarray}
&&
\omega_{12} - \omega_{21} + \omega_{13} - \omega_{31}
 = \alpha (1-x_1) - \alpha H x_1 = \alpha H \left( \frac{1}{H} - x_1 \right)
 \equiv \mu^* N^* \cdot m^* (Q_1 - x_1)
\notag \\ &&
\omega_{21} + \omega_{31} = 
\alpha (H-1) x_1 + r^* N^* x_1 (x_2+x_3) \approx \mu^* N^* \cdot x_1 (1-x_1).
\notag \\ &&
\omega_{12}+\omega_{13}
= \alpha (1-x_1) + r^* N^*x_1(1-x_1) \approx r^* N^*x_1(1-x_1)
\end{eqnarray}

\end{document}  
